**Crystallization of Diblock Copolymer from Microphase Separated Melt**

Chitrita Kundu, Nikhil S. Joshi and Ashok Kumar Dasmahapatra[*]

Department of Chemical Engineering, Indian Institute of Technology Guwahati, Guwahati – 781039, Assam, India

[*]Corresponding author: Phone: +91-361-258-2273; Fax: +91-361-258-2291; Email address: akdm@iitg.ernet.in

## ABSTRACT

Diblock copolymers by virtue of the chemical dissimilarity between the constituting blocks exhibit microphase separation in the melt state. The phase separated melt can successfully be exploited to control the morphology of the final semi crystalline materials by allowing an extended thermal annealing. Thermal annealing accelerates coalescence of microdomains, yielding a phase separated melt that would exhibit a distinctly different crystallization behaviour than microphase separated melt without annealing. In this paper, we report simulation results on the crystallization behaviour of A-B diblock copolymer, wherein the melting temperature of A-block is higher than B-block, instigated from microphase separated melt. During crystallization, the morphological evolution of microphase separated melt is extensively driven by thermal history. Annealing of microphase separated melt at high temperature successfully reorients melt morphology, and remains almost unaltered during the subsequent crystallization (isothermal and non-isothermal), which is attributed to the hard confinement resulted during microphase separation. Annealing induces change in bond orientation of A-block, whereas there is no appreciable change in bond orientation of B-block keeping crystallinity and lamellar thickness unaffected. Isothermal crystallization confines crystallization in phase separated microdomain whereas non-isothermal crystallization results in morphological perturbation of melt microdomain. The rate of crystallization of annealed melt is much faster than the non-annealed melt due to less entanglement and more relaxed structure of achieved through the process of annealing. At higher composition of B-block, A-block produces thicker crystals, which is attributed to the dilution effect exhibited by B-block. Two-step compared to one-step isothermal crystallization yields thicker crystals with higher crystallinity of A-block, whereas the crystallinity and lamellar thickness of the B-block remains same for both the melts.

## 1. INTRODUCTION

The confinement-induced crystallization in block copolymer has accelerated the recent development of nanotechnology as the intrinsic properties and final morphology of a crystal can successfully be tailored by judicious adjustment of constituent block.[1-5] The self-assembled microdomain characteristic of diblock copolymers is one of the convenient ways to achieve nanoscale confinement during crystallization.[5-12] Diblock copolymer consists of two distinct repeat units which are in most of the cases thermodynamically incompatible.[13] This mutual incompatibility leads to microphase separation between blocks offering a large variety of morphologies including lamellar, hexagonally packed cylinder or body centred cubic phases that are stable over a wide range of composition.[14, 15] Microphase separation in bulk provides ordered nanostructures, which are advantageous for designing new functional materials for potential applications such as, in lithography, catalysis, filtration, etc.[16-18]

The crystallization behaviour of amorphous-crystalline diblock copolymer has been considered as a prospective research topic during last few decades.[1, 6, 19-31] Typically, crystallization happens after microphase separation. The microphase separation induces confinement, which restricts crystallization within the respective micro domains, keeping melt morphology intact.[1, 10, 11, 19, 20, 22, 24, 26, 27, 29-31] However, in some cases, the microphase separated structure of semi crystalline diblock copolymer is completely destroyed by the subsequent crystallization of crystalline block[31] producing various morphological patterns.[9, 20, 22, 24, 25, 32, 33] But the morphological perturbation of phase separated melt is typically driven by thermal history.[9, 19, 20, 22, 24] For example, an asymmetric diblock copolymer of polyethylene-*block*-poly(3-methyl-1-butene) exhibits hexagonally packed cylindrical morphology in microphase separated melt. Faster cooling (cooling rate 10-20°C/min) confines crystallization within cylindrical microdomains; whereas slower cooling (cooling rate 8°C/min) ensures morphological perturbation by producing lamellar morphology.[24]

Additionally, the crystallization behaviour of crystalline-crystalline diblock copolymer instigated by microphase separated melt also presents interesting morphological evolution.[5, 7, 12, 34-36] The microdomains can successfully be deployed as potential templates for producing long-range order structures within a polymer matrix. Thus, the incorporation of two different types of crystalline blocks in diblock copolymer offers an effective way to explore polymer crystallization under confinement.[12] During crystallization from microphase separated melt, polyethylene-*block*-poly(ε-caprolactone) (PE-*b*-PCL) diblock copolymer first exhibits an alternate lamellar structure of crystalline PE block and amorphous PCL block, and subsequently, the PCL block crystallizes at a lower crystallization temperatures ($T_c$). When $T_c < 30°C$, the lamellar morphology of PE block remains intact after crystallization of PCL block, whereas at high crystallization temperature ($45°C > T_c > 30°C$), a morphological transition is observed, where PE crystals are fragmentally dispersed in PCL lamellar morphology.[7, 34, 35] The crystallization process of Poly (L-lactide)-*block*-polyethylene (PLLA-*b*-PE) is confined within strongly segregated lamellar microdomain with a path-dependent (viz., one- and two-step cooling) crystallization behaviour. In the first step of the two-step crystallization process (cooling from 190°C to 130°C), PLLA crystallizes first without morphological perturbation of melt microdomain, followed by the crystallization of PE block at 97°C (in the second step cooling from 130 °C to 97 °C). In one-step crystallization from 190°C to 80°C, PE crystallizes at a much faster rate and dictates the final crystal morphology.[5] The orientation and nanostructures of semiconducting polymers play a pivotal role in determining performances of electronic and optoelectronic devices.[36] For example, the crystallization of a conjugated diblock copolymer of poly (2,5-dihexyloxy-p-phenylene)-*block*-(3-hexythiophene) is mainly driven by the crystallization of P3HT, which establishes the final crystal morphology of the thin films. Higher block composition of P3HT promotes breakout crystallization, whereas lower block composition results in confined crystallization.[36]

Similarly, asymmetric syndiotactic polypropylene-*block*-poly(ε-caprolactone) diblock copolymer exhibits hexagonally packed cylindrical morphology in melt, which is completely disrupted, irrespective of the crystallization process, resulting crystalline lamellar morphology. However, in two-step crystallization, interactive crystallization is observed whereas in one-step crystallization confined crystallization is prominent.[12]

Usually, the first crystallizing block suppresses the crystallization of second block during crystallization.[5] However, in some cases, both the blocks crystallize together (viz., coincident crystallization) even if their melting points are widely different, or one block accelerates the crystallization of the other.[3, 37] For example, PPDX-*b*-PCL diblock copolymer exhibits coincident crystallization although there is a significant difference in the melting points of PPDX (100°C) and PCL block (57°C).[3] Recently, Monte Carlo simulation on lattice polymer also reveals that the crystallization of one block accelerates the crystallization of other block.[37]

In our previous work, we have investigated the effect block asymmetry on the crystallization of double crystalline A-B diblock copolymer crystallized from a homogenous melt.[38] In the weak segregation limit, the transition points and the development of crystallinity are extensively governed by the block asymmetry. In contrast, the development of crystallinity and morphological evolution in strong segregation limit are regulated by confinement effect rather than the block asymmetry.[38]

In our present study, we demonstrate the effect of thermal annealing in addition to the block asymmetry on the crystallization of double crystalline diblock copolymer originated by microphase separated melt. Microphase separation introduces self-assembled nanostructures, which can successfully be modified by annealing. The effect of annealing on the subsequent crystallization and morphological development is the main focus of the present study. Annealing at high temperature melt helps to erase the thermal history and influences the overall crystallization process as well as the morphology of the final semi-crystalline structure. The stability of microdomain structure depends on the process of annealing. In amorphous-crystalline diblock copolymer, the coalescence of microdomain structures can be prohibited during crystallization at a low temperature.[39-41] However, if the sample is annealed at a high temperature, microdomains coalescence is possible.[40-42] In addition to this, annealing leads to a change in orientation from edge-on to flat-on lamellar structure.[43-44] The crystallization of all conjugated diblock copolymer of poly(2,5-dihexyloxy-p-phenylene)–block–(3-hexylthiophene) PPP-*b*-P3HT exhibits different morphological behaviour under thermal annealing. When diblock copolymer with higher P3HT is annealed at high temperature, the crystallization breaks out the microphase separated structure. However, the crystallization of diblock copolymer with higher PPP content is confined in microphase separated domains upon annealing.[36] Similarly, thin films of microphase separated poly(butadiene-*block*-ethyleneoxide) diblock copolymer (PB-*b*-PEO) are prepared by spin-coating on silicon wafers. The hydroxyl groups on the surface of the Si wafer interact strongly with PEO and favoured strong adsorption. However, annealing of thin films in molten state leads to a pseudo dewetting forming holes on monolayer.[45-46] In our study, we observe that annealing of microphase separated melt results morphological reorientation while keeping melt morphology intact irrespective of the crystallization processes. However, microphase separated melt without annealing leads to morphological perturbation during non-isothermal crystallization, whereas in isothermal crystallization melt morphology remains intact.

We organize our paper as follows. We discuss modelling and simulation technique in section 2 followed by results and discussions in section 3 and conclusion in section 4.

## 2. MODELLING AND SIMULATION TECHNIQUE

We apply dynamic Monte Carlo (DMC) simulation method to simulate crystallization of diblock copolymer, which has been successfully applied to investigate phase transition of bulk polymers.[37, 38, 47, 48] In our simulation, a polymer chain is represented by joining the successive sites in a simple cubic lattice with a size of $32 \times 32 \times 32$. An initial configuration consists of 480 polymer chains, each with 64 repeat units, which are placed successively one by one in the lattice box, ensuring the connectivity of the chain. Thus, the lattice occupation density is as high as 0.9375, representing a bulk polymer system. The degree of polymerization of a chain is $N$ (viz., 64) with $N_A$ and $N_B$ numbers of A- and B-type repeat units respectively. We express the composition of A-block as $x_A$ ( $N_A / N$ ) and the composition of B-block as $x_B$ ( $N_B / N$ ). A well equilibrated structure is generated by applying a set of microrelaxation algorithms. The microrelaxation algorithm involves a set of Monte Carlo moves such as bond fluctuation, end bond rotation and slithering diffusion.[37, 38, 40, 47, 48] To give further details, we start our simulation by selecting a vacant site randomly from the available vacant sites and then search for a nearest neighbour site occupied either by A- or B-type unit. Appropriate micro relaxation moves are selected in accordance with the position of monomers along the chain. If the selected monomer is terminal one, then end bond rotation and slithering diffusion is implemented with equal probability. On the other hand, if the unit is non-terminal, then single site bond fluctuation move is implemented.[40, 49]

The interaction between A-type and B-type is modelled as the repulsive interaction to represent their mutual immiscibility. The energy retribution to create A-B contact is modelled by $U_{AB}$. The crystallization driving force is modelled as an attractive interaction between neighbouring parallel bonds, and collinear bonds within A- or B- type units and represented by $U_p$ and $U_c$ respectively. The change in energy per Monte Carlo (MC) move is then:

$$\Delta E = -\left(\Delta N_P U_P + \Delta N_c U_c\right)_A - \left(\Delta N_P U_P + \Delta N_c U_c\right)_B + \Delta N_{AB} U_{AB} \tag{1}$$

Where, $\Delta N_p$ and $\Delta N_c$ represents the net change in the number of parallel and collinear bond respectively, for the A- and B- block, and $\Delta N_{AB}$ represents the change in the number of contacts between A- and B-units.

As the block copolymer consists of two different blocks with different melting points, we model B-block as the low melting one and less competent towards crystallization compared to A-block. To implement this, we consider $U_{pB} = \lambda_m U_{pA}$ and $U_{cB} = \lambda_m U_{cA}$. We set $\lambda_m = 0.75$ ($<1$) to represent less driving force for crystallization of B-block compared to A-block. Further, we assume that, $U_p = U_c$ to represent coarse grained interactions in our simulation. In terms of Flory' $\chi$ parameter, segregation strength is calculated as ($\chi N$). The value of ($\chi N$) of our sample system is $(q-2) \times U_{AB} \times N$, where $q$ is the coordination number of our lattice model, $N$ is the degree of polymerization and $U_{AB}$ is demixing energy between two blocks.[40] In our simulation, $U_{AB}$ is calculated as $\lambda U_p$, where $\lambda$ represents segregation strength which is equivalent to Flory's $\chi$ parameter.[38] Thus, the smaller value of $\lambda$ represents weak segregation within our system. In our simulation, we take $\lambda = 1$ to implement weak segregation strength between the blocks. All the energies are normalized by $k_B T$, where, $k_B$ is the Boltzmann constant and $T$ is temperature in Kelvin. Thus, $U_p \sim 1/T$. Now, the change in energy per MC move is modified as follows:

$$\Delta E = \left[ -\left(\Delta N_p + \Delta N_c\right)_A - \lambda_m \left(\Delta N_p + \Delta N_c\right)_B + \lambda \Delta N_{AB} \right] U_p \tag{2}$$

We use the Metropolis sampling scheme with periodic boundary conditions to sample new conformations. The probability of a Monte Carlo move is given by $\exp(-\Delta E)$. We accept new conformation if $\exp(-\Delta E) \geq r$, where $r$ is the random number in the range (0, 1), generated by using the random number generator MT19937.[50] To equilibrate the system, we calculate mean square radius of gyration $\langle R_g^2 \rangle$ as a function of Monte Carlo Steps ($MCS$) (see Figure S1, Supplementary information,[51] for $x_B = 0.5$). We do not observe an appreciable change in the value of $\langle R_g^2 \rangle$ beyond 5000 $MCS$ and it is considered as the equilibration time. We calculate thermodynamics and structural parameters averaged over subsequent 5000 $MCS$.

To monitor crystallization, we calculate fractional crystallinity, $X_c$ of A- and B-block as a function of $U_p$. We define crystallinity ($X_c$) as the ratio of number crystalline bonds to the total number of bonds present in the system. A bond is defined as crystalline if it is surrounded by more than 5 nearest non-bonded parallel bonds. Additionally, we calculate specific heat ($C_v$) as a function of $U_p$. Specific heat ($C_v$) is calculated as equilibrium specific heat from the total energy fluctuations (for all the monomer and comonomer units in simulation box).[38, 47, 48] We calculate average crystallite size $\langle S \rangle$ and lamellar thickness $\langle l \rangle$ as a function of $U_p$ for structural analysis. A crystallite is defined as a small microscopic aggregate having crystalline bonds in same orientation. The crystallite size is defined as the total number of crystalline bonds present in it. We express lamellar thickness $\langle l \rangle$ as the average number of monomer units in the direction of crystal thickness in a given crystallite, and average thickness is calculated over all crystallites present in the system.[38,47,48] We also analyse the orientation of crystalline bonds by calculating bond order parameter ($P$) which is defined as:

$$P = \frac{3\langle \cos^2 \theta \rangle - 1}{2} \qquad (3)$$

Where, $\theta$ is the angle of a concerned bond with reference to Z-axis and <….> represents an assembled average over all the bonds containing more than 10 nearest parallel bonds.[40] According to the definition, if all concerned bonds are in parallel with Z-axis, $P$ is equal to 1, whereas if they are perpendicular with Z-axis, $P$ is equal to -0.5. However, if all concerned bonds are randomly oriented, $P$ is close to zero.[40]

## 3. RESULTS AND DISCUSSIONS

To simulate crystallization of diblock copolymer initiated by microphase separated melt, first we prepare a set of phase segregated melt morphology of various block compositions. Following this, we crystallize the microphase separated melt through non-isothermal and isothermal process.

### 3.1. Preparation of Microphase Separated Melt

Finding a precise location of microphase separation point is one of the important and challenging tasks in our simulation. Microphase separation creates self-assembled microdomain structures which offer spatial confinement within the system during subsequent crystallization.

To drive crystallization, we have considered three potential energies such as, parallel bond interaction energy, collinear bond interaction energy and demixing energy between A- and B-units (Equation 1 and 2), strength of which is given by $\lambda$. This demixing energy promotes phase segregation via microphase separation before crystallization. Therefore, to develop phase separated melt system, we consider only demixing energy and the change in energy per MC move is now modified as:

$$\Delta E = \Delta N_{AB} U_{AB} \qquad (4)$$

We simulate our system from $U_p = 0$ to $U_p = 0.6$ with a step size of 0.02 for different block compositions ($x_B$) ranging from 0.125 to 0.875 with an increment of 0.125. We estimate $C_{v-AB}$ (viz., $C_v$ for A-B contacts) vs. $U_p$ for all block compositions. $C_{v-AB}$ gives a peak as fluctuations in energy and the $U_p$ value associated with the peak is considered as microphase separation point ($U_p^\#$). Figure 1a shows the change in $C_{v-AB}$ with $U_p$ for symmetric block composition (viz., $x_B = 0.5$). The changes in $C_{v-AB}$ with $U_p$ for other compositions are given in Figure S2 of supplementary information.[51] Figure 1b summarizes the change in microphase separation point ($U_p^\#$) with block compositions ($x_B$). For most of the compositions, $U_p^\# = 0.04$, except highly asymmetric diblock copolymer with $x_B = 0.125$ and 0.875. For, $x_B = 0.125$ and 0.875, $U_p^\# = 0.06$ and 0.08, respectively. From the above data, it appears that with the increase in block asymmetry, microphase separation takes place at a relatively lower temperature (viz., higher $U_p$). The above observation comply with poly(L-lactide)-*block*-poly(ε-caprolactone) (PLLA-*b*-PCL) diblock copolymer, which exhibits microphase separation temperature (measured in terms of $T_{ODT}$) at 175 and 220 °C for the sample having 37.4 and 46 wt% PCL block respectively.[52] The snapshot of microphase separated melt morphology for block composition, $x_B = 0.50$ at $U_p = 0.04$ is presented in Figure 2a, where blue and orange line represents A- and B-block units, respectively. The snapshots of remaining compositions are available in Figure S3 of supplementary information.[51]

As discussed before, to generate a phase separated melt, first we simulate our system for $10^4$ $MCS$, out of which 5000 $MCS$ is needed to equilibrate the system, as evident from $\langle R_g^2 \rangle$ vs. $MCS$ trend (Figure S1). After that, we anneal melt morphology for $10^6$ $MCS$ for all the compositions at their respective microphase separation point $U_p^\#$, to allow the chain molecules to relax further, disentangled and generate a better phase separated melt microstructure. The annealed temperature is much higher than the respective melting points of individual blocks. The similar process has been followed by Hong et al. for semi crystalline diblock copolymer of polyethylene-*block*-atactic polypropylene, where the samples are melt annealed at 150°C which is higher than the melting temperature of polyethylene ($T_m \sim 120°C$).[39] Similarly, in asymmetric polyethylene oxide-*block*-poly (1,4-butadiene), the melt sample is prepared by annealing at 80°C for 5 minutes.[42] The snapshot of annealed melt morphology for block composition ($x_B$) of 0.50 is given in Figure 2b. The remaining snapshots are available in Figure S4 of supplementary information.[51]

### 3.2. Non-isothermal Crystallization

We cool our sample system from the respective microphase separation point ($U_p^\#$) to $U_p = 0.6$ with a step size of 0.02 to implement non-isothermal crystallization process. For block composition $x_B = 0.125$, we start simulation from $U_p = 0.06$ and for block composition $x_B = 0.875$, we start simulation from $U_p = 0.08$. For the rest of the compositions, we start our simulation from $U_p = 0.04$.

### 3.2.1. Development of Crystallinity

We monitor crystallization by calculating crystallinity of diblock copolymer originated from microphase separated melt without annealing as well as with annealing. Overall crystallinity ( $X_c$ ) is calculated as the weighted average of the summation of A-block ( $X_A$ ) and B-block ( $X_B$ ): $X_c = x_A X_A + x_B X_B$. The change in overall crystallinity ( $X_c$ ) with $U_p$ introduced by microphase separated melt without annealing and with annealing is available in Figure 3. In Figure 3, there is an abrupt increase in crystallinity at a certain value of $U_p$ and finally reaches to a saturation crystallinity ( $X^{sat}$ ) at $U_p \sim 0.5$ in both the cases (annealed and without annealed samples). The comparison in saturation crystallinity ( $X^{sat}$ ) of both the blocks induced from two different microphase separated melts with block compositions is given in Figure 4. It appears that there is no significant difference in the saturation crystallinity ( $X_A^{sat}$ and $X_B^{sat}$ ) between microphase separated melt without annealing and with annealing. This happens because the development of crystallinity is primarily driven by the degree of cooling. For both type of melts, we use almost same degree of cooling to implement non-isothermal crystallization.

The saturation crystallinity of A-block remains similar with block composition (Figure 4a). However, the saturation crystallinity of B-block ( $X_B^{sat}$ ) shows an increasing trend with increasing block composition (Figure 4b). The enhanced number of B-units at higher $x_B$ facilitates in producing crystalline materials with higher crystallinity. This observation is in accord with the experimental results on the crystallization of poly(ε-caprolactone)-*block*-polyethylene (PCL-*b*-PE) diblock copolymers, wherein the crystallinity of PE blocks increases with the increase of PE content in PCL-*b*-PE diblock copolymer.[35]

### 3.2.2. Bond Orientation

In order to follow the orientation of crystalline bonds with respect to Z-axis, we calculate bond order parameter ($P$) of individual blocks over all the compositions. Figure 5 represents the change in bond orientation of individual blocks for symmetric diblock copolymer viz., ($x_B = 0.50$) from microphase separated melt with and without annealing. The snapshots of asymmetric block copolymer induced by microphase separated melt (viz., $x_B = 0.25$ and 0.75) are available in Figure S5, supplementary information.[51] From the above figures, it is evident that, annealed melt induces change in orientation of A-block compared to microphase separated melt without annealing; whereas there is no appreciable change in B-block during annealing. Annealing produces random orientation for the crystalline bonds of B-block. It happens because microphase separated melt (without annealing) induces morphological perturbation during non-isothermal crystallization. Therefore, the rearrangement of crystalline bonds is possible and it gives perpendicular orientation with respect to Z-axis. However, annealed melt retains the melt morphology set during the annealing, no appreciable rearrangement of crystalline bonds is observed, and the orientation becomes parallel to Z-axis. The confinement induced by crystallization of A-block makes B-block less facile for the re-arrangement of crystalline bonds. We have observed a similar trend in bond orientation for asymmetric diblock copolymers (see Figure S6, Supporting information,[51] for $x_B = 0.25$ and 0.75).

### 3.2.3. Structural Analysis

We calculate average lamellar thickness $\langle l \rangle$ separately for both the blocks as a function of $U_p$ for all compositions ($x_B$). Figure S6 and S7 (Supporting information[51]) represent the variation of $\langle l \rangle$ as a function of $U_p$ induced from microphase separated melt without annealing and with annealing, respectively. The trend of $\langle l \rangle$ vs. $U_p$ is similar to that of crystallinity (Figure 3). We compare the lamellar thickness $\langle l \rangle$ of both the blocks at $U_p = 0.6$ (viz., saturated value) induced from microphase separated melt without annealing and with annealing in Figure 6. There is no remarkable difference in the value of lamellar thickness between two different melt systems (viz., with and without annealing) as we have seen in crystallinity trend.

However, the lamellar thickness, $\langle l \rangle$ of both the blocks shows a non-intuitive behaviour with block composition ($x_B$). Lamellar thickness of A-block, $\langle l_A \rangle$, remains almost constant up to $x_B = 0.5$, due to the confinement effect induced by microphase separated melt microdomains resulting thinner crystals within the system (Figure 6a). The magnitude of $\langle l_A \rangle$ shows a steep increase with increasing $x_B$ beyond 0.50. This increase in value of $\langle l_A \rangle$ at higher value of $x_B$ (viz., lower value of $x_A$) is attributed to the dilution effect exhibited by B block. At higher value of $x_B$, B-block acts like a "solvent", weakens the topological restriction to facilitate the crystallization of A-block, producing thicker crystals.[38] For a better understanding, we calculate the mobility of the polymer chains in terms of mean square displacement of center of mass ($d_{cm}^2$) of both the blocks for two types of melt (with and without annealing). We observe a significant increase in the chain mobility (measured in terms of $d_{cm}^2$) of A-block compared to B-block at high block composition, during crystallization of A-block (viz., at $U_p = 0.3$), as shown in Figure 7. At high block composition, A-block crystallizes within the matrix of B-block, which is still in a molten state and acts like a "solvent", even though they are partially segregated. As a result, instead of confinement effect, B-block imposes less hindrance towards the diffusion of A-block units, facilitating in growing thicker crystals. This observation is in accord with the experimental results of PLLA-*b*-PCL diblock copolymer where at lower compositions of PLLA, PCL (major constituent) acts as a diluent and causes the depression in crystallization temperature.[53] Similar type of dilution effect on PLLA block has been also reported for PLLA-*b*-PEO diblock copolymer.[54] On the other hand, the lamellar thickness of B-block, $\langle l_B \rangle$ does not change effectively with block composition ($x_B$) due to the confinement created by microphase separated melt (Figure 6b).

### 3.2.4. Radius of Gyration

We also compare the change in mean square radius of gyration $\langle R_g^2 \rangle$ with $U_p$ for all the block compositions investigated in both the systems (viz., without and with annealing). Figure 8a shows the change in $\langle R_g^2 \rangle$ with $U_p$ for $x_B = 0.5$ (results for other compositions are available in Figure S8, Supporting information[51]). It is clearly visible from the above figure that the system induced from phase separated melt without annealing gives appreciable change in $\langle R_g^2 \rangle$ value compared to the system induced from annealed melt. When we anneal our microphase separated system for long enough time ($1 \times 10^6 \, MCS$), it yields a more relaxed structure with relatively larger microdomains. Therefore, the change in $\langle R_g^2 \rangle$ is negligible in crystals crystallized from an annealed melt compared to that of without annealing. By definition, the mean square radius of gyration is the average squared distance of any unit from the center of mass of a polymer chain. This is an important parameter to understand morphological evolution. We find a significant change in the value of radius of gyration of melt without annealing during crystallization which triggers morphological perturbation (Figure 9a). On the other hand, for annealed melt there is no significant change in the value of radius of gyration, which leads to an unperturbed morphology (Figure 9b).

To compare the change in $\langle R_g^2 \rangle$ in annealed and without annealed melt across all the composition, we calculate the ratio of $\langle R_g^2 \rangle$ for both the melts (viz., $\langle R_g^2 \rangle_{without} / \langle R_g^2 \rangle_{with}$), and plotted as a function of block compositions, $x_B$ (Figure 8b). Figure 8b clearly demonstrates that the change in $\langle R_g^2 \rangle_{without} / \langle R_g^2 \rangle_{with}$ is relatively less in highly asymmetric diblock copolymer (viz., $x_B = 0.125$ and $0.875$) compared to the rest of the compositions. Figure 9a displays the snapshot of semi-crystalline structure (at $U_p = 0.6$) of crystals for $x_B = 0.50$, crystallized from microphase separated melt without annealing. Snapshots of rest of the composition are available in Figure S9, Supporting information.[51] We observe a morphological perturbation of phase separated melt during crystallization over all the compositions except highly asymmetric diblock copolymer ($x_B = 0.125$ and $0.875$). This happens because microphase separated melt without annealing is associated with more intra- and inter-chain entanglement, and relatively less relaxed structure, which produces melt microdomain that can be modified during non-isothermal crystallization.

On the other hand, the morphology of microphase separated melt with annealing remains almost unperturbed during crystallization. Annealing of microphase separated melt develops microdomains in melt morphology which is less facile to be modified upon crystallization, irrespective of the block compositions. The snapshots of semi-crystalline structure crystallized from annealed melt for $x_B = 0.50$ is available in Figure 9b. Snapshots of the rest of the compositions are available in Figure S10, Supporting information.[51]

### 3.3. Isothermal Crystallization

To execute isothermal crystallization, we quench our sample system from the respective microphase separation point, $U_p^{\#}$ (see Figure 1b) to $U_p = 0.6$ directly and annealed for $10^5$ $MCS$.

### 3.3.1. Development of Crystallinity

We observe the development of crystallinity with Monte Carlo Steps crystallized from microphase separated melt without annealing and with annealing, for all compositions. We also calculate scaled overall crystallinity (Figure 10), $X_c^* = (X_c - X_{ci}) / (X_{cf} - X_{ci})$, ranges from 0 to 1 for individual blocks as a function $MCS$. $X_{ci}$ and $X_{cf}$ represent the crystallinity at the beginning and the end of the isothermal crystallization process. Change in the scaled crystallinity ($X_c^*$) of individual blocks of diblock copolymer crystallized from microphase separated melt without annealing and with annealing is presented in Figure S11 and S12 (Supporting information), respectively.[51] The trend in isothermal crystallinity reveals that the transition kinetics for two different melts follows a similar pathway. The scaled crystallinity is useful for the calculation of Avrami index, which gives an idea about crystal geometry (viz., two-dimensional or three-dimensional). As the crystallization driving force for isothermal crystallization (viz., $U_p = 0.6$) is sufficient to introduce crystallinity for both the blocks, the mode of crystallization is coincident crystallization, where both the blocks crystallize simultaneously. The above observation is in line with the isothermal crystallization of the phase separated melt of poly (ρ-dioxanone)-block-poly(ε-caprolactone) diblock copolymer, where crystallization kinetics of both the blocks overlap.[3]

### 3.3.2. Radius of Gyration

We compare the change in mean square radius of gyration, $\langle R_g^2 \rangle$ with *MCS* for symmetric diblock copolymer introduced by microphase separated melt without and with annealing in Figure 11a. Results for the rest of the compositions are given in Figure S13, Supporting information.[51] There is no substantial change in the value of $\langle R_g^2 \rangle$ with *MCS* in both the melt systems, however, the magnitude of $\langle R_g^2 \rangle$ of microphase separated annealed melt is higher than that of microphase separated melt without annealing. The morphology set during the microphase separation and subsequent annealing remains unperturbed upon isothermal crystallization. In non-isothermal crystallization, which follows a stepwise cooling method, allows the chain segments to change conformational pattern, and we have observed a gradual increase in $\langle R_g^2 \rangle$ upon cooling for without annealed melt (see Figure 8a). However, in isothermal crystallization, where the sample is directly quenched to $U_p = 0.6$ from the respective $U_p^{\#}$, the conformational change is restricted due to the onset of crystallization. We plot the saturation value of radius gyration at $U_p = 0.6$ for all the block compositions (with and without annealing) in Figure 11b. The magnitude of $\langle R_g^2 \rangle$ of microphase separated annealed melt is higher compared to the microphase melt without annealing for most of the block compositions except for highly asymmetric block. The non-monotonic trend in $\langle R_g^2 \rangle$ with composition is attributed to the block asymmetry present in the system. Figure 12a and 12b represent the snapshots of final crystal structure (viz., at $U_p = 0.6$) of symmetric diblock copolymer isothermally crystallized from microphase melt without annealing and with annealing, respectively. Snapshots of the rest of the compositions for without and with annealing are available in Figure S14 and S15 (Supporting information), respectively.[51] These snapshots clearly demonstrate that the molecular arrangement of phase separated melt morphology is almost remain unperturbed during isothermal crystallization, wherein the development of crystallinity is fast enough to restrict a morphological rearrangement (viz., perturbation).

### 3.3.3. Avrami Index

Time evolution of crystallinity can be described by Avrami equation[55]:

$$(1 - X_c^*) = \exp(-kt^n) \qquad (5)$$

Where $X_c^*$ represents the scaled crystallinity ranges from 0 to 1, $n$ is the Avrami index, indicative of crystal geometry. We estimate the value of Avrami index $(n)$ based on the primary crystallization,[56] for both the blocks as a function of block composition (Figure 13). The value of Avrami index $(n)$ for both the blocks ranges from 0.5 to 1.1, which indicates the formation of two-dimensional crystals via homogeneous nucleation. Lower value of Avrami index $(n)$ attributes to the restricted crystal growth under confinement due to microphase separated melt morphology. This result is in line with the partially miscible poly (L-lactide)-block-poly (ε-caprolactone) diblock copolymer, which follows a homogeneous nucleation pathway showing the Avrami index ~ 1.0.[53] During isothermal crystallization of PCL block in asymmetric PLLA-*b*-PCL diblock copolymer, PCL block exhibits first order transition kinetics with Avrami Index close to 1.0.[57, 58] Similarly, crystallization of crystalline-amorphous diblock copolymers where crystalline block is confined within the microdomains of amorphous block also follows homogeneous nucleation mechanism[2, 26, 59] with Avrami index ~ 1.0.[26, 59, 60] For example, crystallization of polyethylene oxide, confined within a large number of microdomains of polystyrene exhibits homogeneous nucleation.[2] Similarly, PLLA block in PLLA-*b*-PS diblock copolymer also follows a homogeneous nucleation with Avrami index close to $1.0.^{26}$ However, the value of Avrami Index $(n)$ is smaller for both the blocks crystallized from microphase melt without annealing compared to that of annealed melt. The system with more relaxed microdomain structure gives a relatively higher value of Avrami Index $(n)$ due to less entanglement effect in the polymer matrix.

### 3.3.4. Crystallization Half-time

We calculate crystallization half-time ($t_{1/2}$) in terms of the number of $MCS$, to get an approximate idea of the rate of crystallization. We estimate $t_{1/2}$ as the number of $MCS$ required to have crystallinity equal to 50% of the saturated value (viz., at the end of isothermal crystallization). Table 1 displays the value of $t_{1/2}$ for both the blocks, in terms of $MCS$ for all the compositions of diblock copolymer crystallized from microphase separated melt with and without annealing. From the above table, we observe that annealed melt crystallizes at a relatively faster rate in comparison with the microphase separated melt without annealing. This change in rate with annealing happens due to the presence of less entanglement and more relaxed structure of the annealed melt, which generates stable micro domains. On the other hand, microphase separated melt without annealing would need to relax the melt structure over a few more $MCS$ before the crystallization. Thus, the rate of crystallization is much faster for annealed melt than melt without annealing.

### 3.4. Two-step Isothermal Crystallization

We perform two-step isothermal crystallization to examine the effect of quench depth on the crystallization of microphase separated melt. In the first step, we cool the equilibrated system from the respective microphase separation point ($U_p^\#$) to $U_p = 0.3$ and annealed for $10^5$

Monte Carlo Steps. In this process, only A-block is crystallized while B-block remains in a molten state. Following this, we quench the system from $U_p = 0.3$ to $U_p = 0.6$ and annealed for $10^5 \, MCS$ to initiate crystallization of B-block. We compare saturation crystallinity ( $X^{sat}$ ) of both the blocks during two-step and one-step isothermal crystallization of diblock copolymer crystallized from microphase separated melt, without annealing as well as with annealing in Table 2 and 3, respectively.

The process of annealing in microphase separated melt is introduced by the implementation of $10^6$ Monte Carlo Steps at the respective microphase separation points. We observe a comparable increase of crystallinity of A-block in two-step crystallization than that of one-step crystallization; whereas, the crystallinity of B-block remains almost same for both the processes. The changing mode of crystallization is responsible for this significant difference in crystallinity. Two-step cooling follows a sequential crystallization mechanism, where the development of crystallinity of A-block is unaffected by the crystallization of B-block. One-step cooling follows a coincident crystallization mechanism, where both the blocks experience a competition for crystallization. As a result, crystallization (also crystallinity) of A-block is hindered. Similarly, we observe that the lamellar thickness of A-block is higher in two-step compared to one-step isothermal crystallization (Table S1 and S2, Supporting information)[51] for both the melts.

The above observation is in accord with Hoffman-Weeks formula, which describes the development of crystallinity majorly governed by degree of cooling.[60] When we implement two-step cooling at $U_p = 0.3$ (first step), the crystallization driving force induces crystallization of A-block whereas at $U_p = 0.6$ (second step), the crystallization driving force induces crystallization of B-block. Due to the difference in degree of cooling, A-block produces different lamellar thickness in two-step compared to one-step isothermal crystallization.

However, in both the processes, the degree of cooling for the B-block is similar as it is crystallized at $U_p = 0.6$. Therefore, the crystallinity and lamellar thickness is same for B-block irrespective of two-step and one-step isothermal crystallization. The snapshots of symmetric diblock copolymer (viz., $x_B = 0.50$) introduced by microphase separated melt without and with annealing for two-step isothermal crystallization are shown in Figure 14. The snapshots of asymmetric diblock copolymer introduced by microphase separated melt without and with annealing (viz., $x_B = 0.25$ and $0.75$) are available in Figure S16 and S17 (Supporting information), respectively.[51] There is no morphological perturbation of phase separated melt morphology in isothermal two-step cooling for symmetric as well as asymmetric block. The above observation is in line with the crystallization behaviour of poly (L-lactide)-block-polyethylene (PLLA-*b*-PE) diblock copolymer, where two-step isothermal crystallization preserves melt morphology intact.[5]

## 4. CONCLUSIONS

The prospect of diblock copolymer has emerged in nanotechnology and biomedical application over few decades.[16, 18, 61] In block copolymer lithography[16] or in organic photovoltaic cells, the self-assembled microdomain characteristics are widely applicable. Simulation study of diblock copolymer crystallization, crystallized from microphase separated melt is reported with two different patterns of melt morphology (viz., annealed and without annealed). We observe a morphological perturbation during non-isothermal crystallization for diblock copolymer crystallized from microphase separated melt without annealing, whereas the melt morphology remains unperturbed for diblock copolymer crystallized from microphase separated annealed melt, as evidenced from the trend of $\left\langle R_g^2 \right\rangle$. Annealing induces re-orientation of chain segments and facilitates the coalescence of microdomains created during microphase separation. The morphology set during the annealing remains unaffected after crystallization (isothermal and non-isothermal). Highly asymmetric diblock copolymer (viz., high composition of the B-block) shows an enhancement of lamellar thickness of A-block, which is attributed to the dilution effect shown by B-block. This dilution effect is observed in both the melts (with and without annealing).

We study isothermal crystallization with two different modes of cooling (viz., one- and two-step) to understand the effect of quench depth on crystallization. We implement one-step isothermal crystallization by quenching microphase phase separated melt directly from the respective microphase separation point ( $U_p^{\#}$ ) to $U_p = 0.6$, which results in denial of morphological perturbation irrespective of melt morphology (viz., annealed and without annealed microphase separated melt). Simultaneously, we execute two–step isothermal crystallization by cooling microphase separated melt from the respective microphase separation point to $U_p = 0.3$, followed by cooling from $U_p = 0.3$ to $U_p = 0.6$. Two-step crystallization yields better crystallinity of A-block compared to one-step isothermal crystallization, but the crystallinity of B-block remains identical for both the melts without morphological change. Crystallization of B-block happens in the presence of the confinement created during the crystallization of A-block. As a result, no morphological change is observed for B-block during crystallization. Our findings suggest that understanding on the morphological development with annealing (cf., varying the annealing temperature) would enable to tune the semi-crystalline morphology of diblock copolymers.

**CONFLICTS OF INTEREST**

There are no conflicts to declare.

## ACKNOWLEDGMENT

N.S.J. acknowledges the financial support from the CSIR, Govt. of India (sanction letter no. 22(0638)/13/EMR-II). Computational facility supported by the SERB, Department of Science and Technology (DST), Government of India (sanction letter no. SR/S3/CE/0069/2010) is highly acknowledged.

## REFERENCES

[1]Y. L. Loo, R. A. Register, A. J. Ryan, and G. T. Dee, Macromolecules **34**, 8968 (2001).

[2]A. J. Muller, V. Balsamo, M. L. Arnal, T. Jakob, H. Schmalz, and V. Abetz, Macromolecules **35**, 3048 (2002).

[3]J. Albuerne, L. Marquez, A. J. Muller, J. M. Raquez, P. Degee, P. Dubois, V. Castelletto, and I. W. Hamley, Macromolecules **36**, 1633 (2003).

[4]D. Shin, K. Shin, K. A. Aamer, G. N. Tew, T. P. Russel, J. H. Lee, and J. Y. Jho, Macromolecules **38**, 104 (2005).

[5]R. V. Castillo, A. J. Muller, M. C. Lin, H. Chen, U. S. Jeng, and M. A. Hillmyer, Macromolecules **41**, 6154 (2008).

[6]Y. L. Loo, R. A. Register, and A. J. Ryan, Phys. Rev. Lett. **84**, 4120 (2000).

[7]S. Nojima, Y. Akutsu, A. Washino, and S. Tanimoto, Polymer **45**, 7317 (2004).

[8]T. Sakurai, Y. Ohguma, and S. Nojima, Polym. J. **40**, 971 (2008).

[9]M. C. Lin, H. L. Chen, W. F. Lin, P. S. Huang, and J. C. Tsai, J. Phys. Chem. B **116**, 12357 (2016).

[10]R. H. Lohwasser, G. Gupta, P. Kohn, M. Sommer, A. S. Lang, T. Thurn Albrecht, and M. Thelakkat, Macromolecules **46**, 4403 (2013).

[11]L. Chen, J. Jiang, L. Wei, X. Wang, G. Xue, and D. Zhou, Macromolecules **48**, 1804 (2015).

[12]M. C. Lin, H. L. Chen, W. B. Su, C. J. Su, U. S. Jeng, F. Y. Tzeng, J. Y. Wu, J. C. Tsai, and T. Hashimoto, Macromolecules **45**, 5114 (2012).

[13]S. B. Darling, Prog. Polym. Sci. **32**, 1152 (2007).

[14]F. S. Bates and G. H. Fredrickson, Annu. Rev. Phys. Chem. **41**, 525 (1990).

[15]F. S. Bates and G. H. Fredrickson, Phys. Today **52**, 32 (1999).

[16]C. M. Bates, M. J. Maher, D. W. Janes, C. J. Ellison, and C. G. Willson, Macromolecules **47**, 2 (2014).

[17]D. Yao, K. Zhang, and Y. Chen, Polymer **94**, 1 (2016).

[18]F. H. Schacher, P. A. Rupar, and I. Manners, Angew. Chem. Int. Ed. **51**, 7898 (2012).

[19]L. Zhu, S. Z. D. Cheng, B. H. Calhoun, Q. Ge, R. P. Quirk, E. L. Thomas, B. S. Hsiao, F. Yeh, and B. Lotz, Polymer **42**, 5829 (2001).

[20]M. Ueda, K. Sakurai, S. Okamoto, D. J. Lohse, W. J. MacKnight, S. Shinkai, S. Sakurai and S. Nomura, Polymer **44**, 6995 (2003).

[21]T. Shiomi, H. Tsukada, H. Takeshita, K. Takenaka, and Y. Tezuka, Polymer **42**, 4997 (2001).

[22]T. Shiomi, H. Takeshita, H. Kawaguchi, M. Nagai, K. Takenaka, and M. Miya, Macromolecules **35**, 8056 (2002).

[23]A. J. Ryan, I. W. Hamley, W. Bras, and F. S. Bates, Macromolecules **28**, 3860 (1995).

[24]D. J. Quiram, R. A. Register, and G. R. Marchand, Macromolecules **30**, 4551 (1997).

[25]S. Nojima, H. Nakano, and T. Ashida, Polymer **34**, 4168 (1993).

[26]R. M. Ho, F. H. Lin, C. C. Tsai, C. C. Lin, B. T. Ko, B. S. Hsiao, and I. Sics, Macromolecules **37**, 5985 (2004).

[27]I. W. Hamley, J. P. A. Fairclough, N. J. Terrill, A. J. Ryan, P. M. Lipic and F. S. Bates, Macromolecules **29**, 8835 (1996).

[28]l. W. Hamley, J. P. A. Fairclough, A. J. Ryan, F. S. Bates and E. Towns Andrews, Polymer **37**, 4425 (1996).

[29]Y. L. Loo, R. A. Register, and A. J. Ryan, Macromolecules **35**, 2365 (2002).

[30]S. Nojima, M. Fujimoto, H. Kakihira, and S. Sasaki, Polym. J. **30**, 968 (1998).

[31]S. Nojima, M. Toei, S. Hara, S. Tanimoto, and S. Sasaki, Polymer **43**, 4087 (2002).

[32]S. Nojima, K. Kato, S. Yamamoto, and T. Ashida, Macromolecules **25**, 2237 (1992).

[33]S. Nojima, H. Nakano, Y. Takahashi, and T. Ashida, Polymer **35**, 3479 (1994).

[34]S. Nojima, Y. Akutsu, M. Akaba, and S. Tanimoto, Polymer **46**, 4060 (2005).

[35]S. Nojima, K. Ito, and H. Ikeda, Polymer **48**, 3607 (2007).

[36]X. Yu, H. Yang, S. Wu, Y. Geng, and Y. Han, Macromolecules **45**, 266 (2012).

[37]Y. Li, Y. Ma, J. Li, X. Jiang, and W. Hu, J. Chem. Phys. **136**, 104906 (2012).

[38]C. Kundu and A. K. Dasmahapatra, J. Chem. Phys. **141**, 044902 (2014).

[39]S. Hong, A. A. Bushelman, W. J. MacKnight, S. P. Gido, D. J. Lohse and L. J. Fetters, Polymer **42**, 5909 (2001).

[40]W. Hu, Macromolecules **38**, 3977 (2005).

[41]L. Zha and W. Hu, Prog. Polym. Sci. **54-55**, 232 (2016).

[42]Y. Y. Huang, C. H. Yang, H.-L. Chen, F.-C. Chiu, T.-L. Lin, and W. Liou, Macromolecules **37**, 486 (2004).

[43]Y. Wang, C. M. Chan, K. M. Ng, and L. Li, Macromolecules **41**, 2548 (2008).

[44]R. E. Prudhomme, Prog. Polym. Sci. **54–55**, 214 (2016).

[45]G. Reiter, G. Castelein, P. Hoerner, G. Riess, A. Blumen and J. U. Sommer, Phys. Rev. Lett. **83**, 3844 (1999).

[46]G. Reiter, J. Polym. Sci. Part B Polym. Phys. **41**, 1869 (2003).

[47]A. K. Dasmahapatra, H. Nanavati, and G. Kumaraswamy, J. Chem. Phys. **131**, 074905 (2009).

[48]C. Kundu and A. K. Dasmahapatra, Polymer **55**, 958 (2014).

[49]W. Hu, V. B. F. Mathot, and D. Frenkel, Macromolecules **36**, 2165 (2003).

[50]T. Nishimura and M. Matsumoto, http://www.math.sci.hiroshima-u.ac.jp/~m-mat/MT/emt.html (2002)

[51]See supplementary material at (URL) for the appendixes which include change in mean square radius of gyration $\left\langle R_g^2 \right\rangle$ with Monte Carlo Steps at $U_p = 0$; Change in specific heat of AB contacts ( $C_{v-AB}$ ) with $U_p$; Snapshots of microphase separated melt without and with annealing at respective microphase separation point ($U_p^\#$); Change in bond order parameter ( $P$ ) with $U_p$ induced from microphase separated melt without annealing and with annealing; Change in average lamellar thickness of A-block $\left\langle l_A \right\rangle$ and B-block $\left\langle l_B \right\rangle$ with $U_p$ for different $x_B$ induced from microphase separated melt without and with annealing; Change in mean square radius of gyration $\left\langle R_g^2 \right\rangle$ with $U_p$ induced from microphase separated melt without and with annealing; Snapshots of semi-crystalline structure of diblock copolymer induced from microphase separated melt without and with annealing at $U_p = 0.6$; Change in scaled crystallinity ( $X_c^*$ ) with Monte Carlo Steps ( $MCS$ ) at $U_p = 0.6$ for A- and B-block introduced by microphase separate melt without annealing during one-step isothermal crystallization.

[52]J. K. Kim, D. J. Park, M. S. Lee, and K. J. Ihn, Polymer **42**, 7429 (2001).

[53]R. V. Castillo, A. J. Muller, J. M. Raquez, and P. Dubois, Macromolecules **43**, 4149 (2010).

[54]J. Sun, Z. Hong, L. Yang, Z. Tang, X. Chen, and X. Jing, Polymer **45**, 5969 (2004).

[55]M. Avrami, J. Chem. Phys. **7**, 1103 (1939).

[56]A. T. Lorenzo, M. L. Arnal, J. Albuerne, and A. J. Müller, Polym. Test. **26**, 222 (2007).

[57]I. W. Hamley, P. Parras, V. Castelletto, R. V. Castillo, A. J. Müller, E. Pollet, P. Dubois, and C. M. Martin, Macromol. Chem. Phy. **207**, 941 (2006).

[58]R. M. Michell and A. J. Müller, Prog. Polym. Sci. **54–55**, 183 (2016).

[59]A. J. Müller, V. Balsamo, and M. L. Arnal, *Block Copolymers II*, (Springer Berlin Heidelberg, Berlin, Heidelberg, 2005).

[60]H. Marand, J. Xu, and S. Srinivas, Macromolecules **31**, 8219 (1998).

[61]A. J. Muller, J. Albuerne, L. Marquez, J.-M. Raquez, P. Degee, P. Dubois, J. Hobbs, and I. W. Hamley, Farad. Discuss. **128**, 231 (2005).

**Table captions:**

**Table 1: Comparison in crystallization half-time ($t_{1/2}$, in terms of number *MCS*) of A and B-block separately for diblock copolymer induced from microphase separated melt without annealing and with annealing.**

**Table 2: Comparison in saturation crystallinity of diblock copolymer induced from microphase separated melt without annealing during two-step and one-step isothermal crystallization.**

**Table 3: Comparison in saturation crystallinity of diblock copolymer induced from microphase separated annealed melt during two-step and one-step isothermal crystallization.**

**Figure captions:**

**Figure 1: (a) Change in specific heat of AB contacts ($C_{v-AB}$) with $U_p$ for $x_B = 0.50$. (b) Change in microphase separation point ($U_p^{\#}$) with $x_B$.**

**Figure 2: Snapshots of microphase separated melt (a) without annealing and (b) with annealing at $U_p = 0.04$ for $x_B = 0.50$.**

**Figure 3: Change in overall crystallinity ( $X_c$ ) with $U_p$ for different $x_B$ induced from microphase separated melt (a) without annealing and (b) with annealing.**

**Figure 4: Comparison in saturation crystallinity of (a) A-block and (b) B-block induced from microphase separated melt without and with annealing.**

**Figure 5: Change in bond order parameter ( $P$ ) with $U_p$ for $x_B$ = 0.50 for (a) A-block and (b) B-block.**

**Figure 6: Comparison in average lamellar thickness of (a) A-block, $\langle l_A \rangle$ and (b) B-block, $\langle l_B \rangle$ induced from microphase separated melt with and without annealing.**

**Figure 7: Change in mean square displacement of centre of mass with block composition for microphase separated melt (a) without annealing and (b) with annealing at $U_p$ = 0.3.**

**Figure 8: Change in (a) mean square radius of gyration, $\langle R_g^2 \rangle$ with $U_p$ for $x_B$ = 0.5, (b) Change in the ratio of mean square radius of gyration of microphase separated melt without annealing and with annealing, with $U_p$ for all the composition ( $x_B$ ).**

**Figure 9: Snapshot of semi crystalline structure of diblock copolymer of $x_B$ = 0.50 induced from microphase separated melt (a) without annealing and (b) with annealing.**

**Figure 10: Change in overall crystallinity ($X_c$) with Monte Carlo Steps (*MCS*) for diblock copolymer introduced by microphase separate melt (a) without annealing and (b) with annealing during one-step isothermal crystallization.**

**Figure 11: Change in (a) mean square radius of gyration, $\langle R_g^2 \rangle$ with Monte Carlo Steps (*MCS*) for $x_B = 0.5$ of microphase separated melt without annealing and with annealing (b) Change in mean square radius of gyration, $\langle R_g^2 \rangle$ with block composition at $U_p = 0.6$.**

**Figure 12: Snapshots of semi crystalline structures of diblock copolymer of $x_B = 0.50$ induced from microphase separated melt (a) without annealing and (b) with annealing during isothermal crystallization.**

**Figure 13: Change in Avrami Index ($n$) with $x_B$ induced from microphase separated melt without annealing and with annealing, for (a) A-block and (b) B-block.**

**Figure 14: Snapshots of semi crystalline structures of symmetric diblock copolymer induced from microphase separated melt (a) without annealing at $U_p = 0.3$, (b) without annealing at $U_p = 0.6$, (c) with annealing at $U_p = 0.3$ and (d) with annealing at $U_p = 0.6$ during two-step isothermal crystallization.**

**Table 1**

| $x_B$ | A-Block | | B-block | |
|---|---|---|---|---|
| | Without | With | Without | With |
| 0.125 | 210 | 180 | 1870 | 1100 |
| 0.25 | 400 | 250 | 1770 | 1165 |
| 0.375 | 470 | 280 | 1275 | 800 |
| 0.50 | 650 | 380 | 1100 | 720 |
| 0.625 | 855 | 465 | 935 | 630 |
| 0.75 | 875 | 400 | 550 | 350 |
| 0.875 | 650 | 460 | 290 | 240 |

**Table 2**

| $x_B$ | A-Block | | B-block | |
|---|---|---|---|---|
| | **Two-step** | **One-step** | **Two-step** | **One-step** |
| **0.125** | **0.73** | **0.65** | **0.54** | **0.56** |
| **0.25** | **0.72** | **0.62** | **0.57** | **0.58** |
| **0.375** | **0.71** | **0.61** | **0.59** | **0.59** |
| **0.50** | **0.71** | **0.60** | **0.59** | **0.60** |
| **0.625** | **0.72** | **0.60** | **0.60** | **0.61** |
| **0.75** | **0.73** | **0.60** | **0.63** | **0.62** |
| **0.875** | **0.76** | **0.62** | **0.65** | **0.66** |

**Table 3**

| $x_B$ | A-Block | | B-block | |
|---|---|---|---|---|
| | **Two-step** | **One-step** | **Two-step** | **One-step** |
| 0.125 | 0.73 | 0.65 | 0.58 | 0.56 |
| 0.25 | 0.72 | 0.63 | 0.58 | 0.60 |
| 0.375 | 0.71 | 0.62 | 0.60 | 0.61 |
| 0.50 | 0.72 | 0.61 | 0.60 | 0.62 |
| 0.625 | 0.72 | 0.61 | 0.62 | 0.63 |
| 0.75 | 0.73 | 0.62 | 0.64 | 0.64 |
| 0.875 | 0.75 | 0.64 | 0.66 | 0.67 |

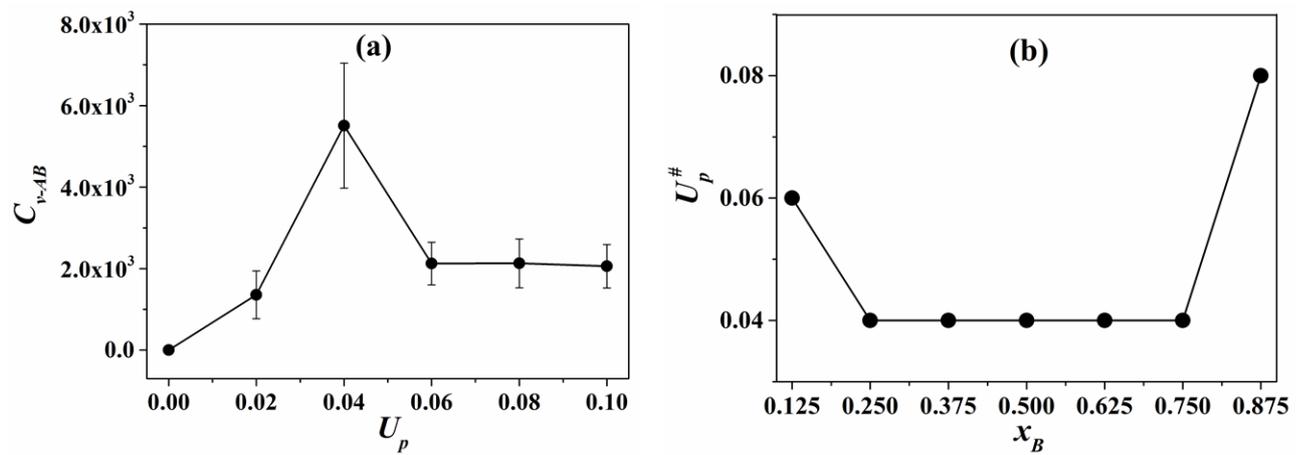

**Figure 1**

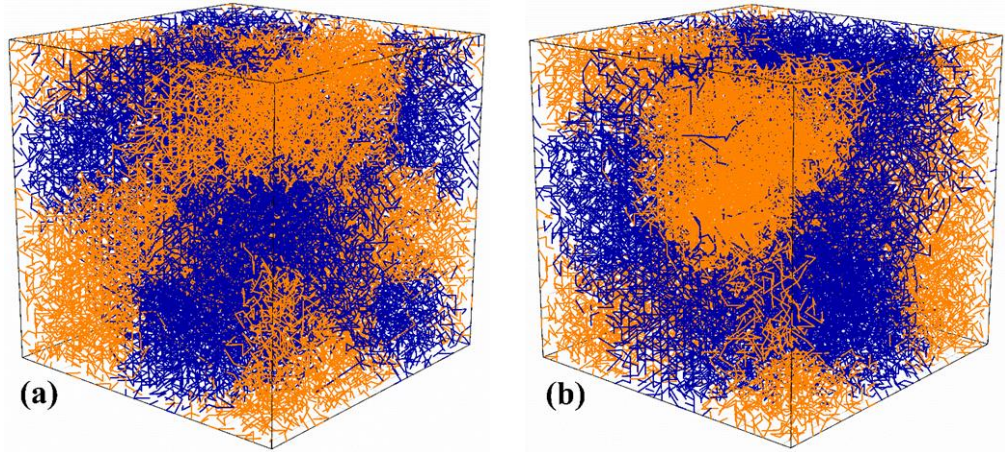

**(a)**  **(b)**

**Figure 2**

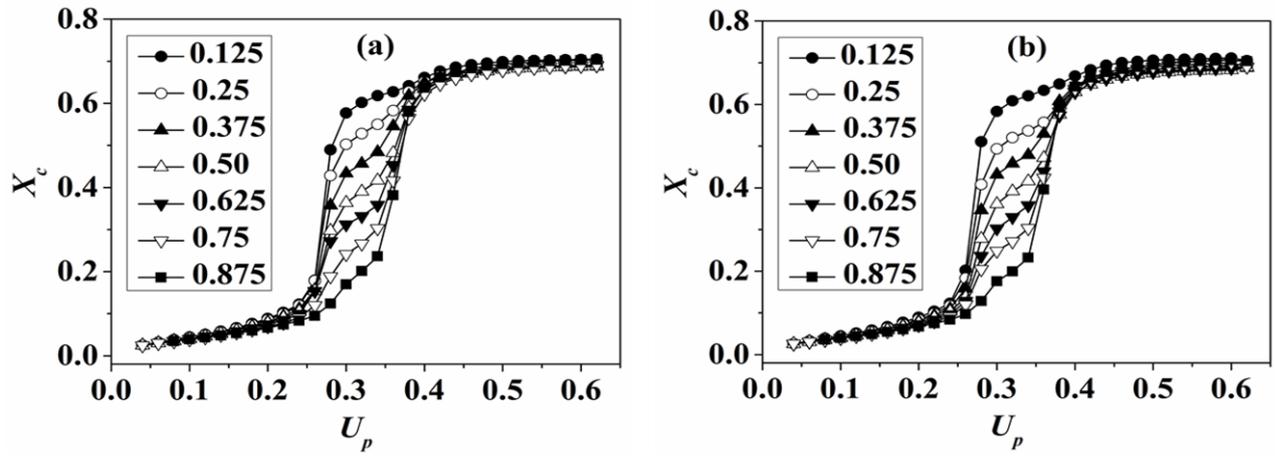

**Figure 3**

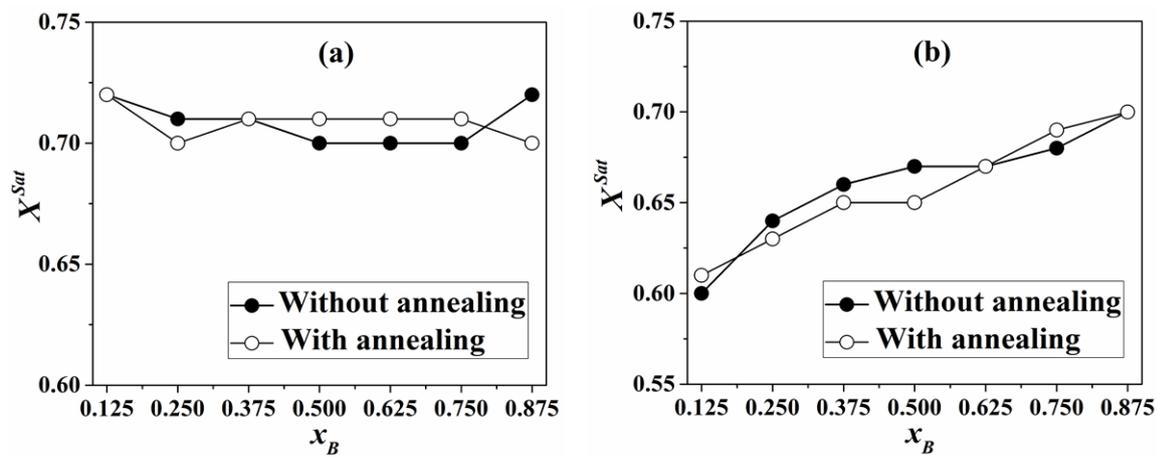

**Figure 4**

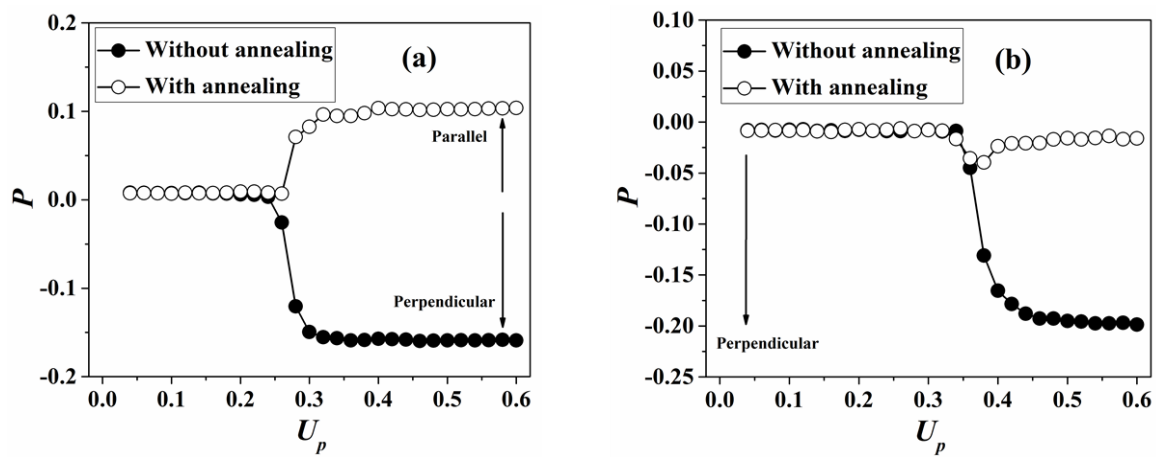

**Figure 5**

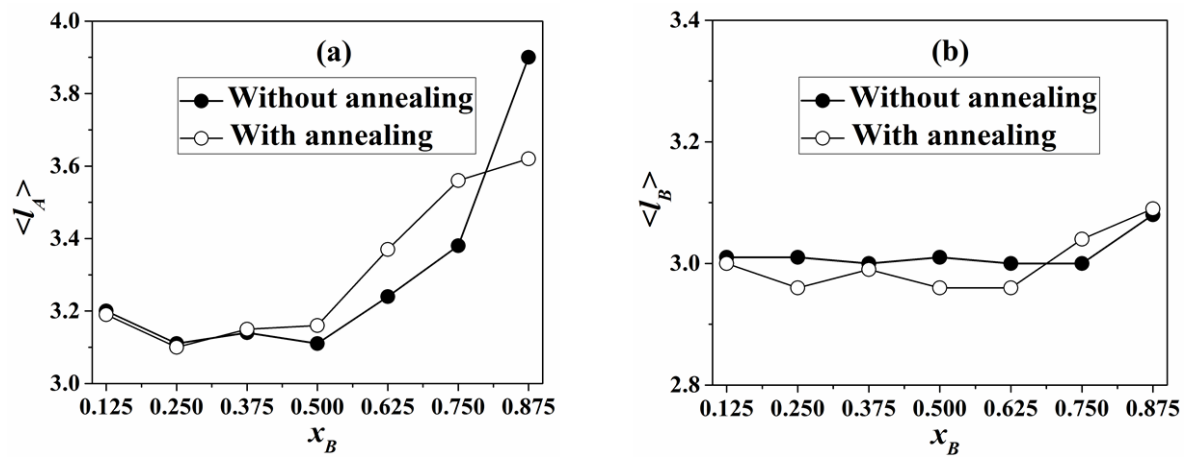

**Figure 6**

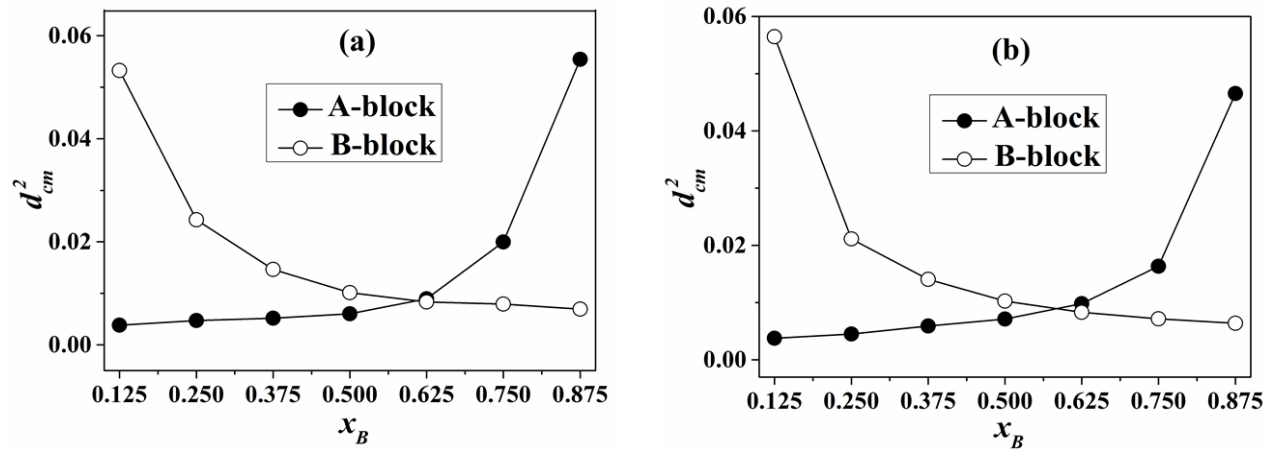

**Figure 7**

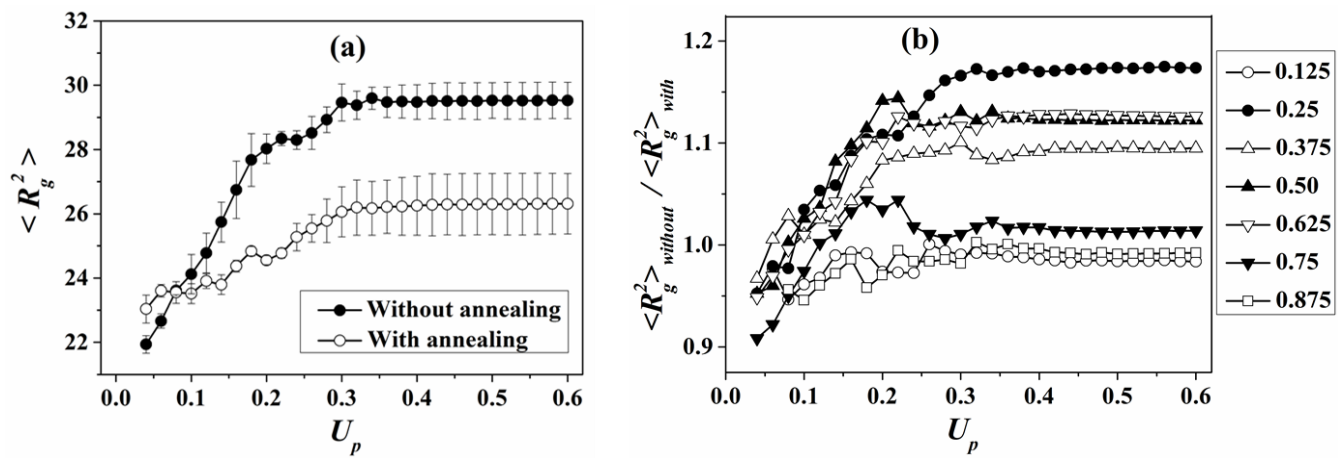

**Figure 8**

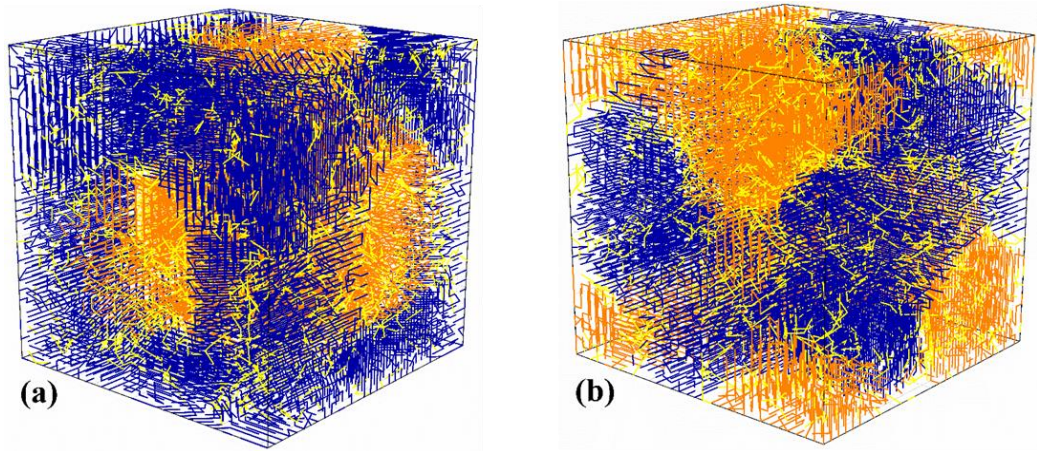

**(a)** **(b)**

**Figure 9**

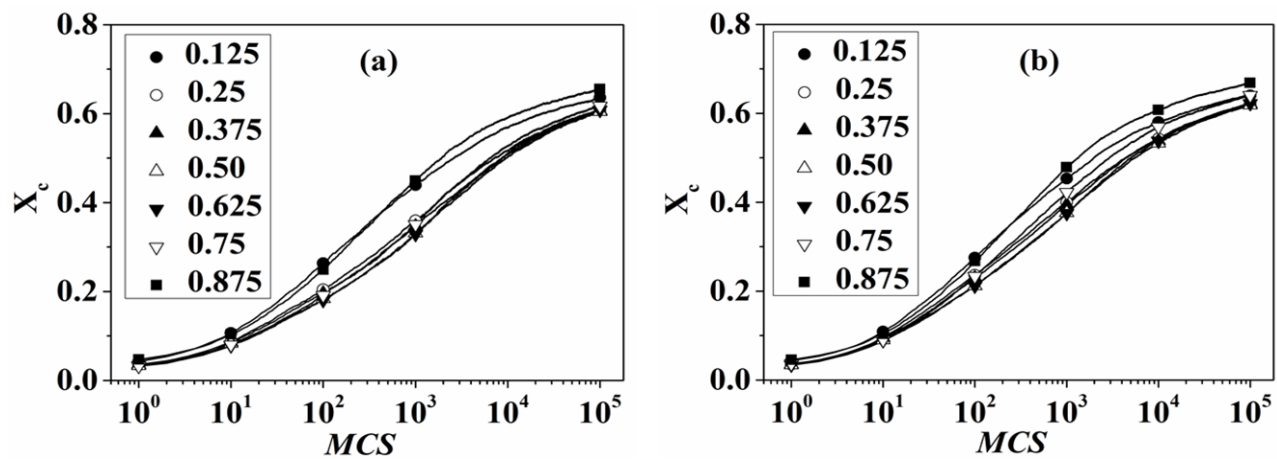

**Figure 10**

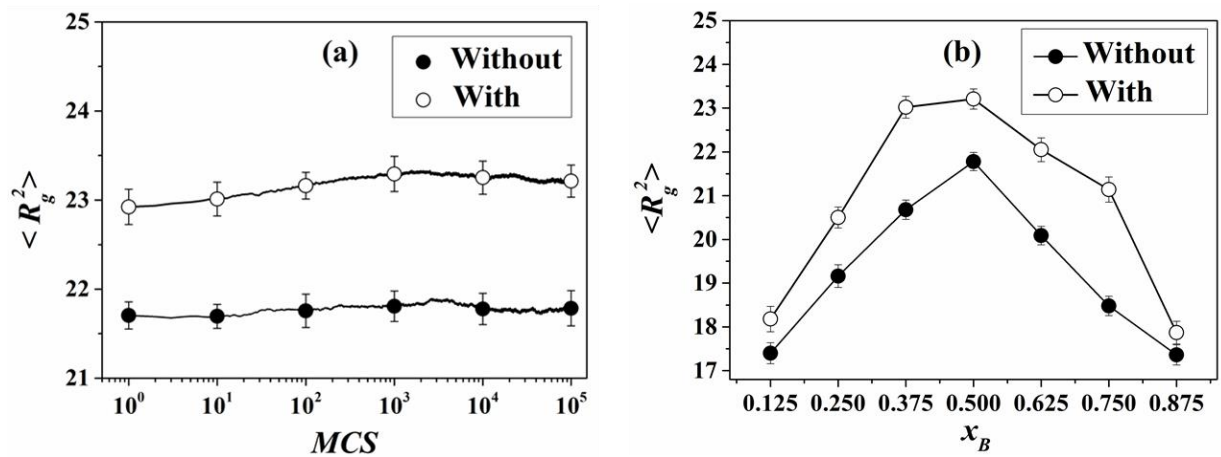

**Figure 11**

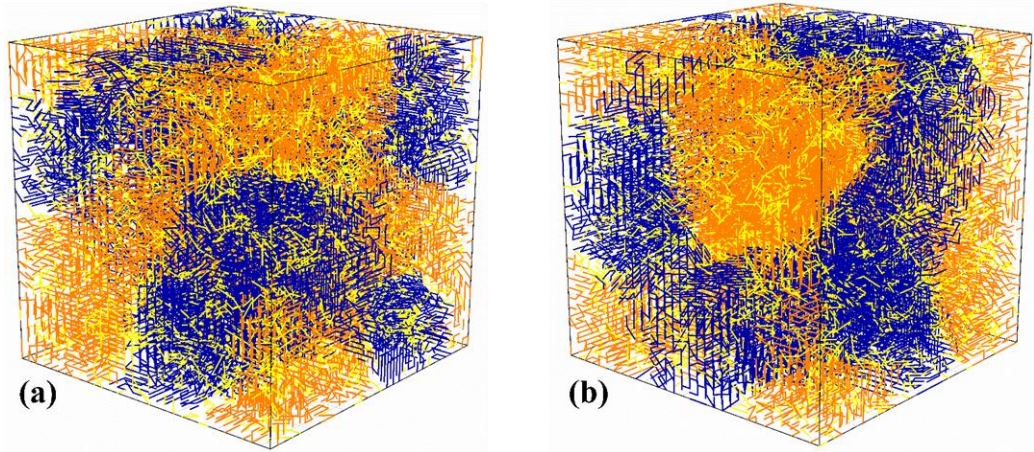

**(a)**　　**(b)**

**Figure 12**

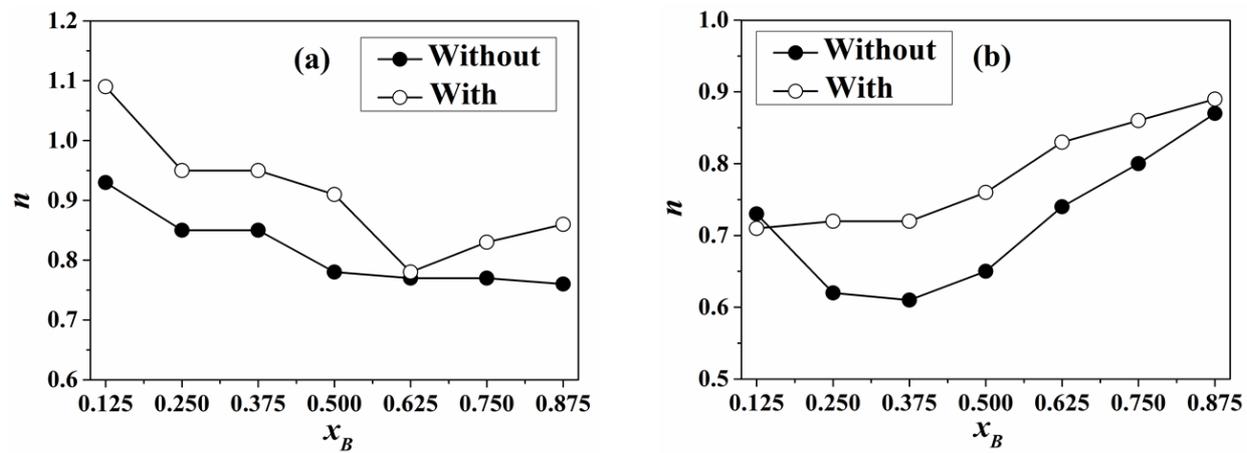

**Figure 13**

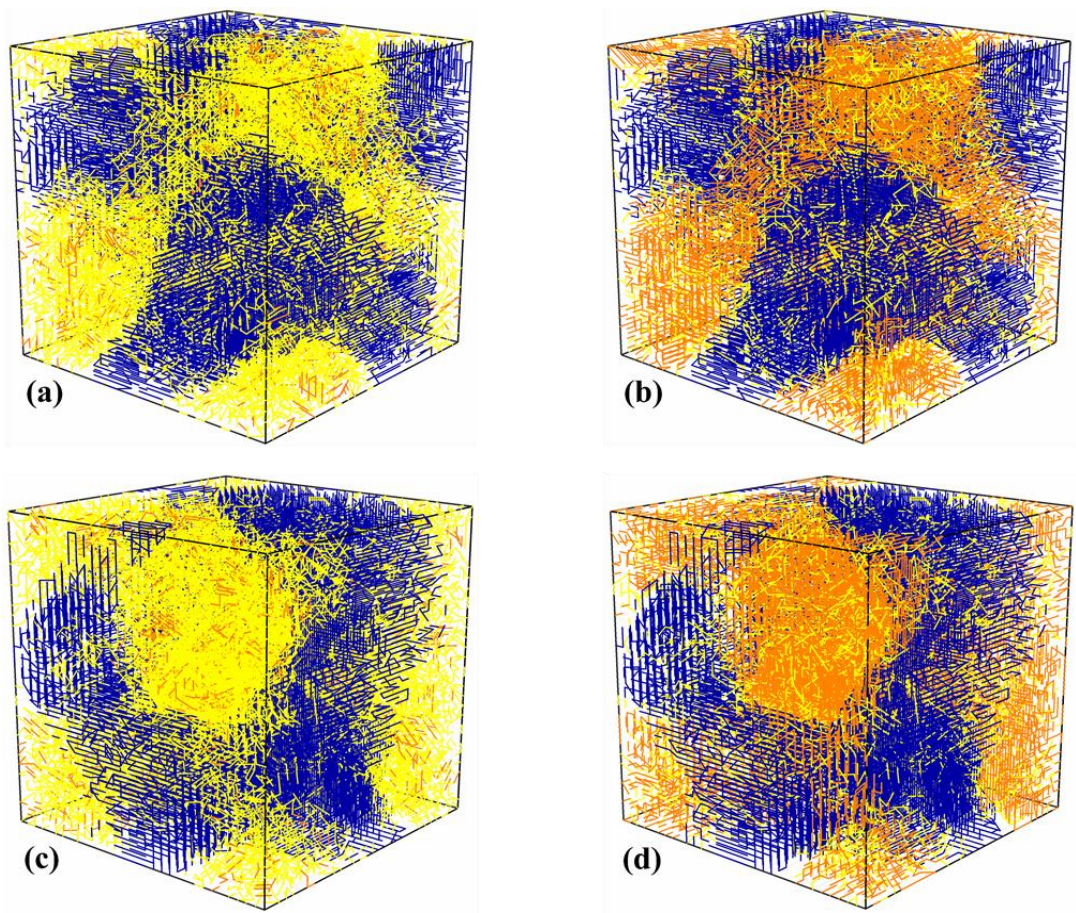

**Figure 14**

# Supporting Information for

## Crystallization of Diblock Copolymer from Microphase Separated Melt

Chitrita Kundu, Nikhil S. Joshi and Ashok Kumar Dasmahapatra[*]

Department of Chemical Engineering, Indian Institute of Technology Guwahati, Guwahati –

781039, Assam, India

**Figure S1. Change in mean square radius of gyration** $\left\langle R_g^2 \right\rangle$ **with Monte Carlo Steps at** $U_p = 0$ **for** $x_B = 0.50$. **There is no appreciable change in the value of** $\left\langle R_g^2 \right\rangle$ **beyond 5000 MCS and it is considered as the equilibration time.**

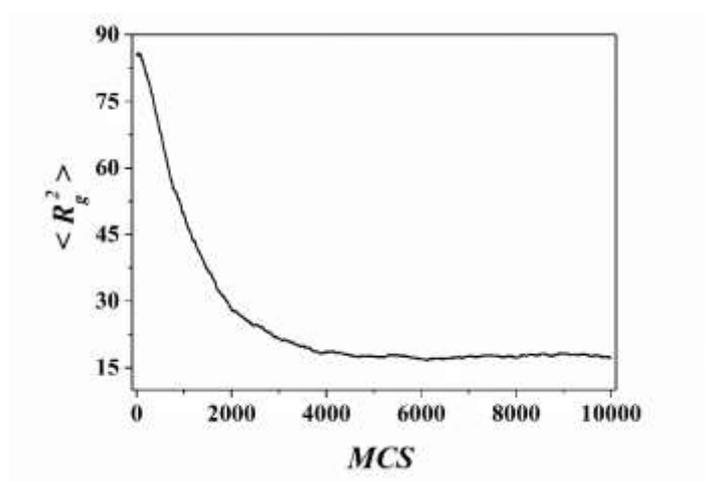

[*] Corresponding author: Phone: +91-361-258-2273; Fax: +91-361-258-2291; Email address: akdm@iitg.ernet.in

**Figure S2. Change in specific heat of AB contacts ($C_{v-AB}$) with $U_p$ for (a) $x_B = 0.125$, (b) $x_B = 0.25$, (c) $x_B = 0.375$, (d) $x_B = 0.625$, (e) $x_B = 0.75$, and (f) $x_B = 0.875$. There is a peak due to fluctuations in demixing energy and the corresponding $U_p$ is considered as microphase separation point ($U_p^{\#}$).**

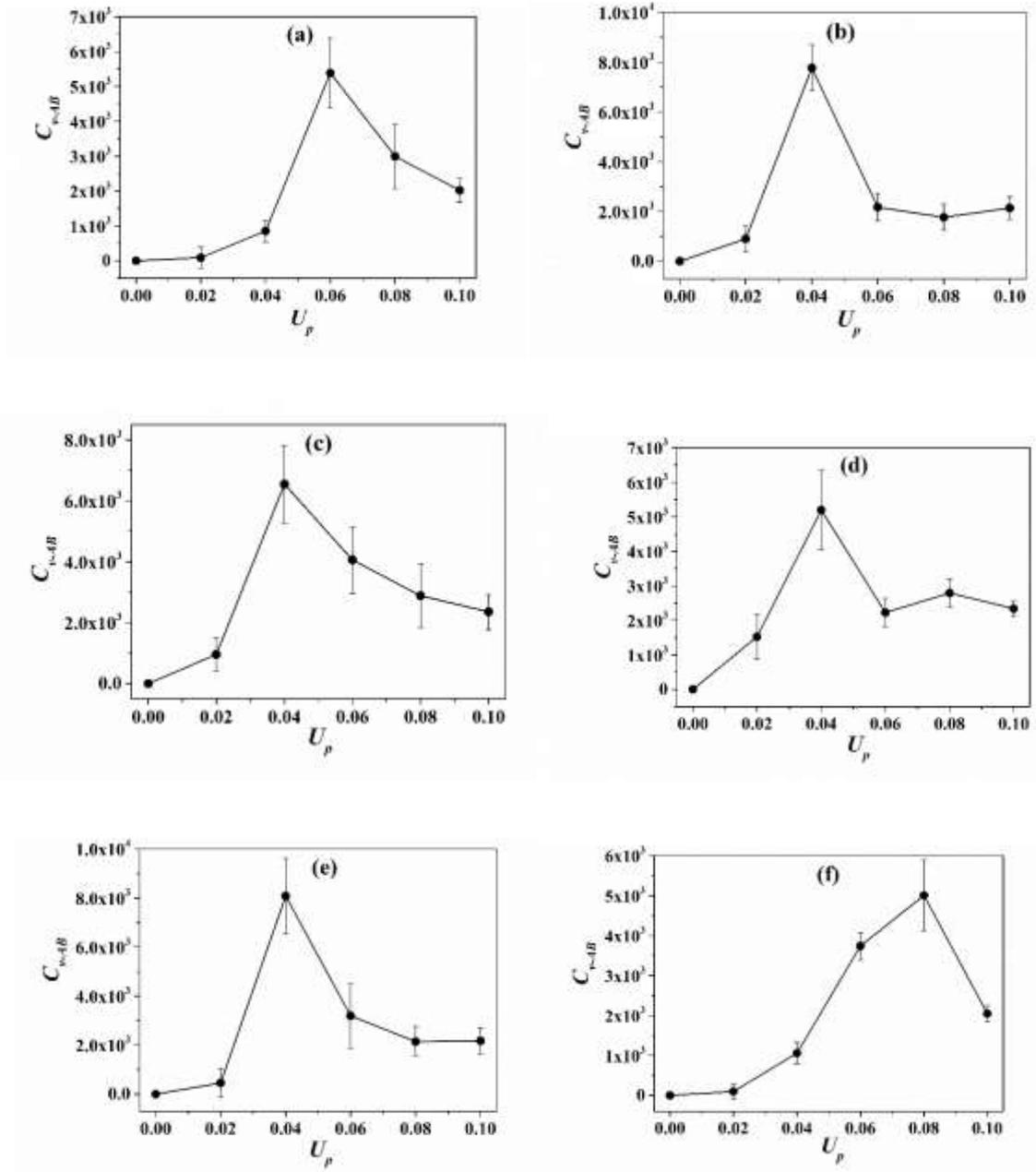

**Figure S3. Snapshots of microphase separated melt without annealing at respective microphase separation point** ($U_p^\#$) **for (a)** $x_B = 0.125$, **(b)** $x_B = 0.25$, **(c)** $x_B = 0.375$, **(d)** $x_B = 0.625$, **(e)** $x_B = 0.75$ **and (f)** $x_B = 0.875$. **Blue and orange lines represent A-block and B-block respectively.**

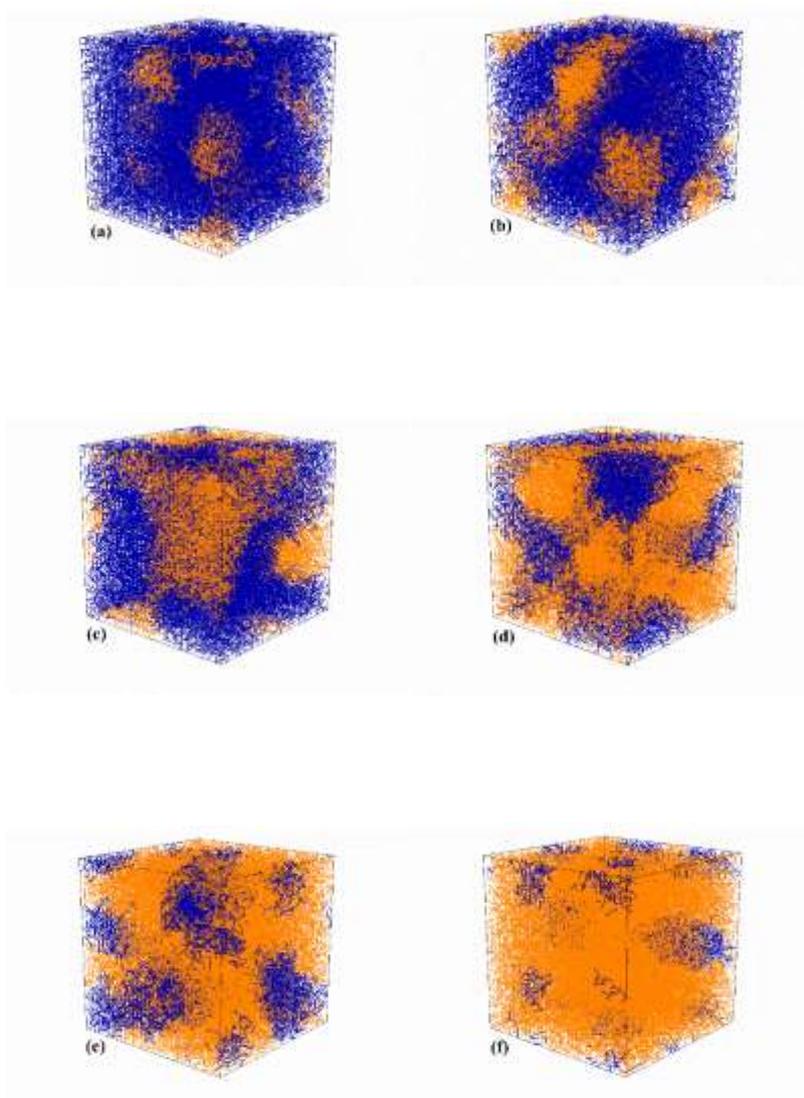

**Figure S4. Snapshots of microphase separated annealed melt at respective microphase separation point** $(U_p^{\#})$ **for (a)** $x_B = 0.125$, **(b)** $x_B = 0.25$, **(c)** $x_B = 0.375$, **(d)** $x_B = 0.625$, **(e)** $x_B = 0.75$ **and (f)** $x_B = 0.875$**. Blue and orange lines represent A-block and B-block respectively.**

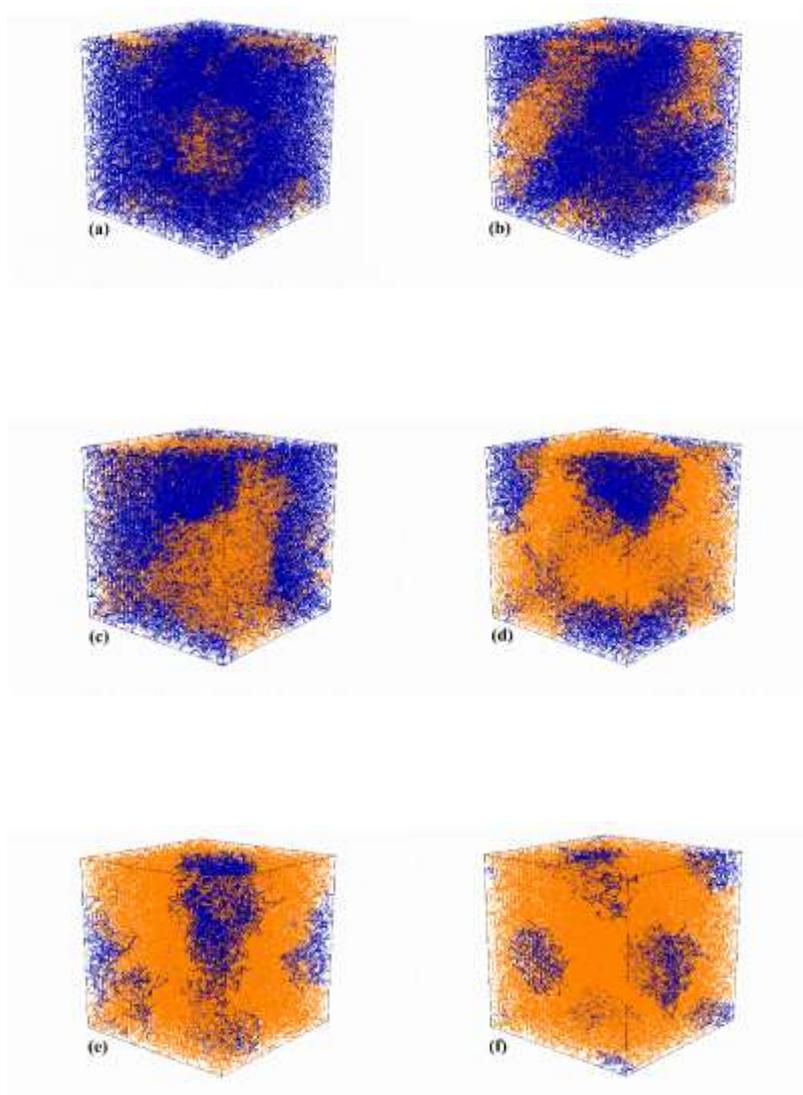

**Figure S5. Change in bond order parameter ( $P$ ) with $U_p$ for (a) A-block of $x_B = 0.25$, (b) B-block of $x_B = 0.25$, (c) A-block of $x_B = 0.75$ and (d) B-block of $x_B = 0.75$ induced from microphase separated melt without annealing and with annealing.**

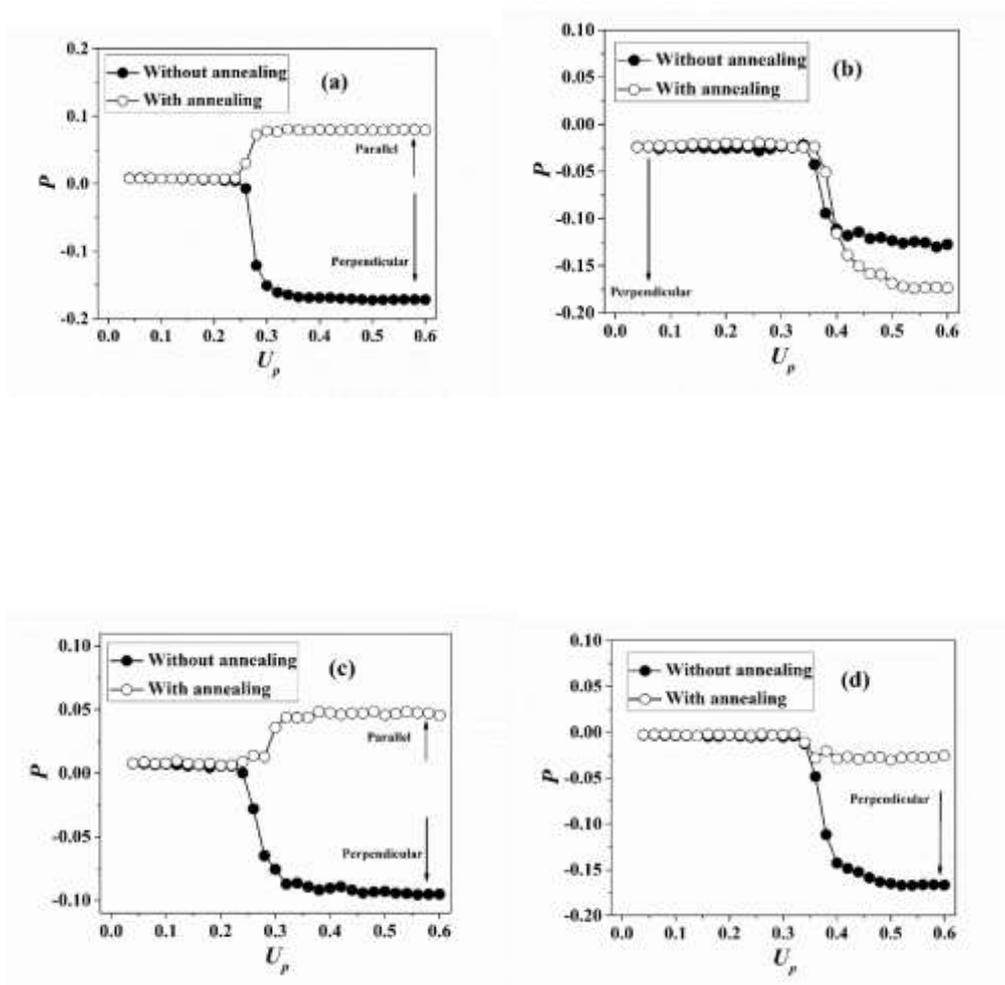

**Figure S6. Change in average lamellar thickness of (a) A-block $\langle l_A \rangle$ and (b) B-block $\langle l_B \rangle$ with $U_p$ for different $x_B$ induced from microphase separated melt without annealing. With the increment of $U_p$, average lamellar thickness $\langle l \rangle$ increases to saturation value.**

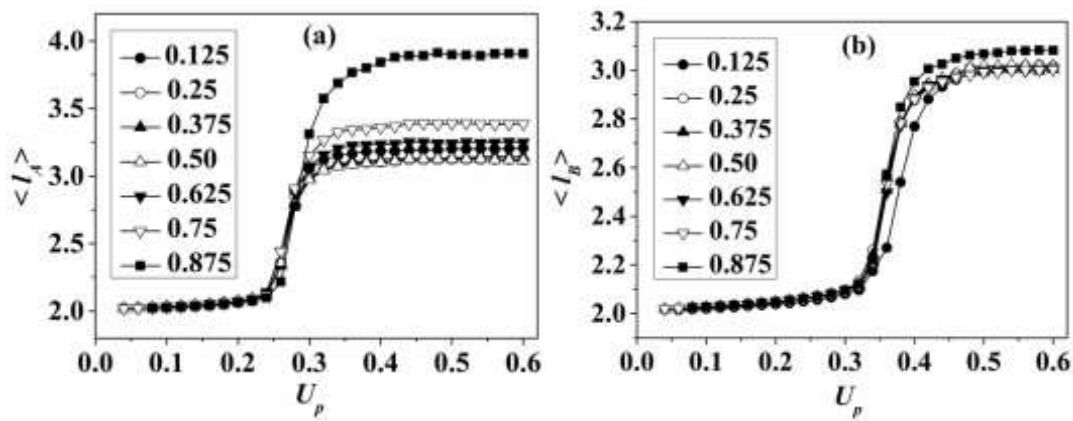

**Figure S7. Change in average lamellar thickness of (a) A-block $\langle l_A \rangle$ and (b) B-block $\langle l_B \rangle$ with $U_p$ for different $x_B$ induced from microphase separated annealed melt. With the increment of $U_p$, average lamellar thickness $\langle l \rangle$ increases to saturation value.**

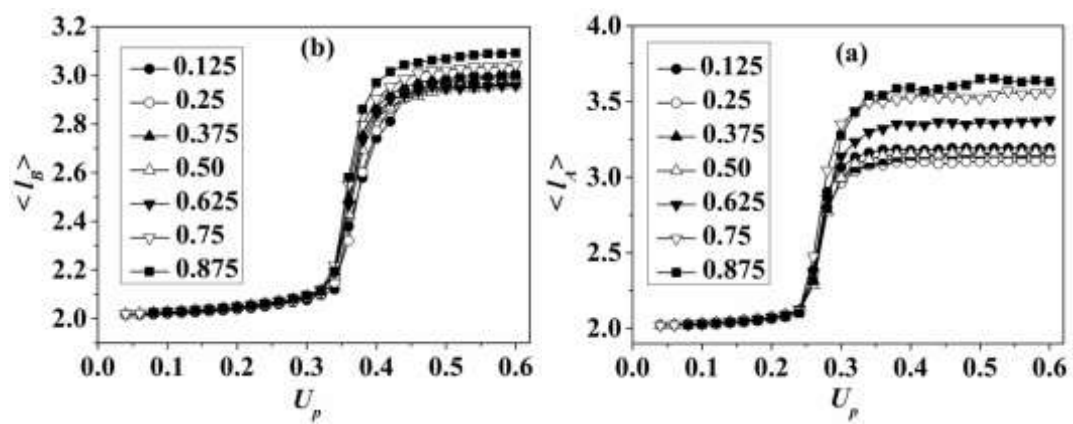

**Figure S8: Change in mean square radius of gyration $\left\langle R_g^2 \right\rangle$ with $U_p$ of (a) $x_B = 0.125$, (b) $x_B = 0.25$, (c) $x_B = 0.375$, (d) $x_B = 0.625$, (e) $x_B = 0.75$ and (f) $x_B = 0.875$ induced from microphase separated melt without and with annealing. There is an appreciable change in the value of $\left\langle R_g^2 \right\rangle$ of diblock copolymer generated from microphase separated melt without annealing compared to with annealing. But the change is negligible for highly asymmetric block (viz., $x_B = 0.125$ and 0.875).**

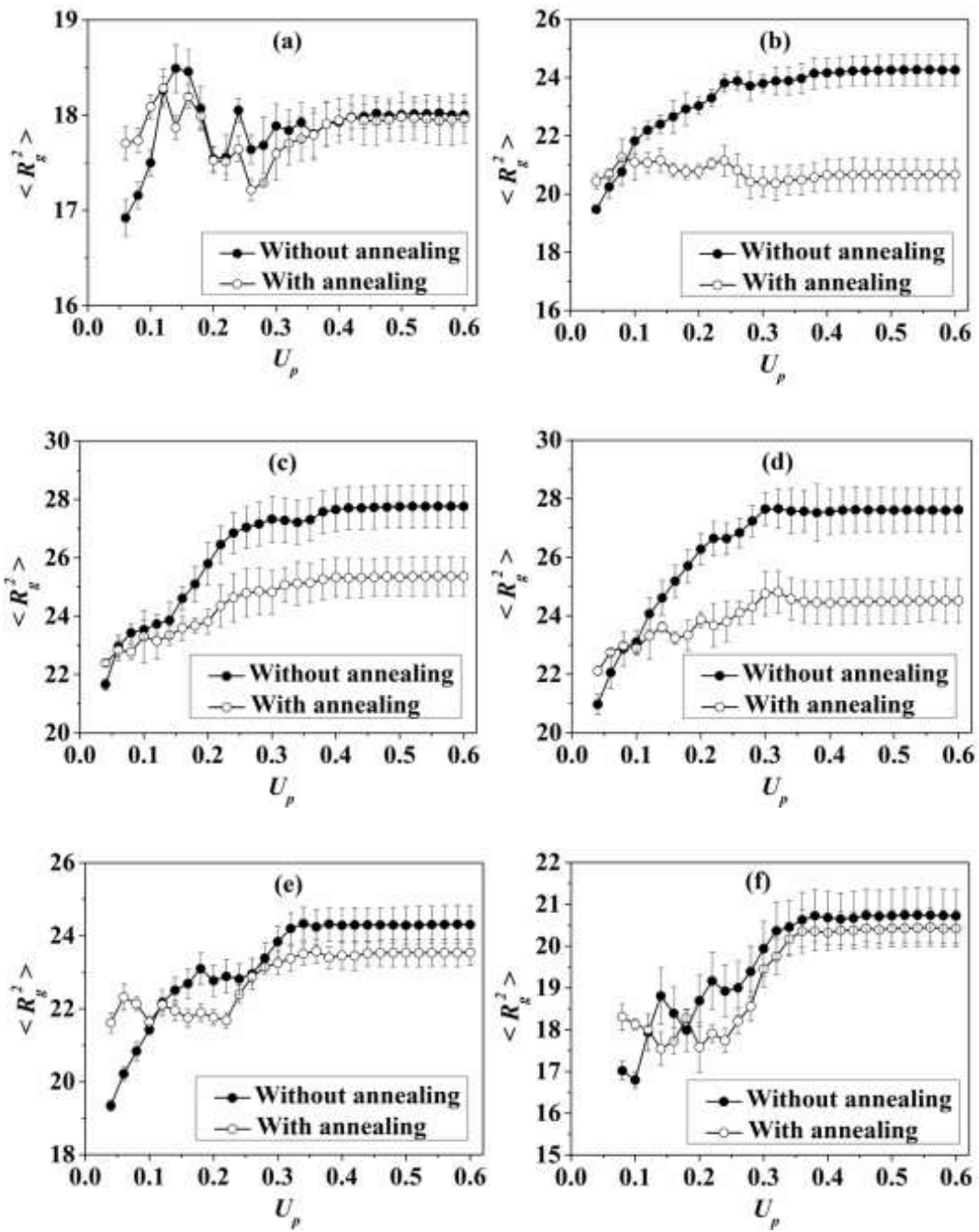

**Figure S9. Snapshots of semi-crystalline structure of diblock copolymer induced from microphase separated melt without annealing at $U_p$ = 0.6 for (a) $x_B$ = 0.125, (b) $x_B$ = 0.25, (c) $x_B$ = 0.375, (d) $x_B$ = 0.625, (e) $x_B$ = 0.75 and (f) $x_B$ = 0.875. Blue and orange lines represent crystalline bonds of A-block and B-block respectively, and yellow lines represent non-crystalline bonds of both the blocks.**

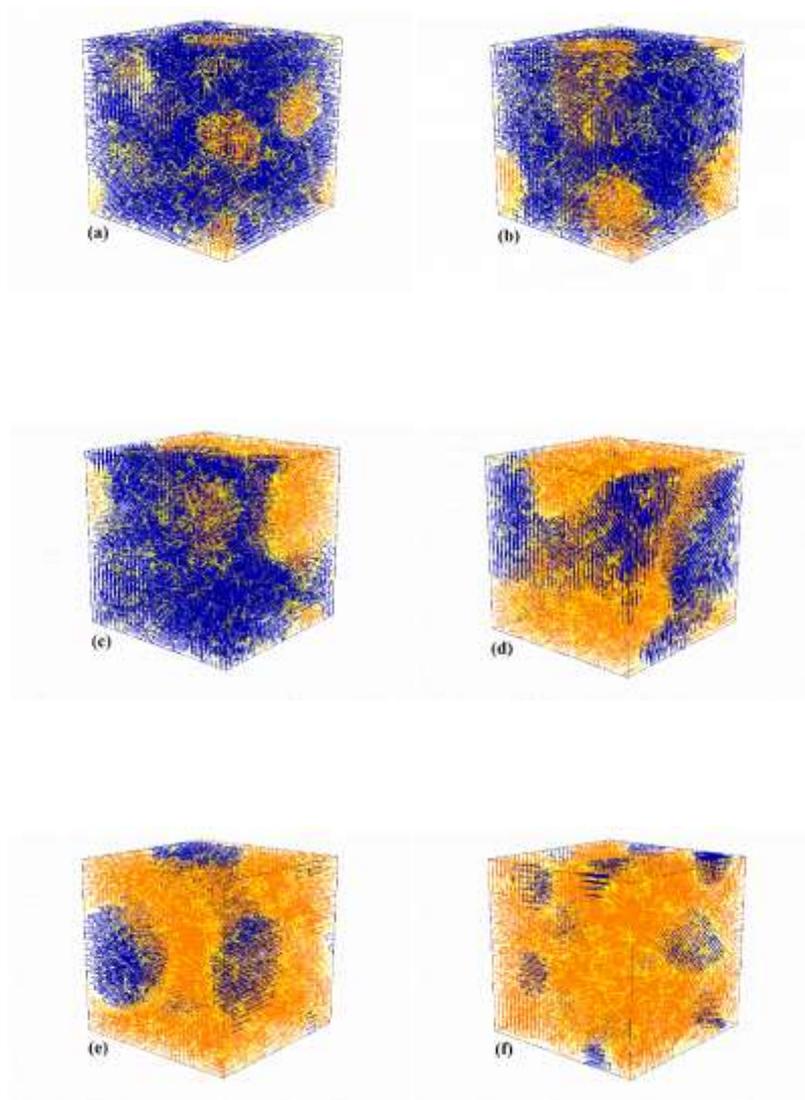

**Figure S10. Snapshots of semi-crystalline structure of diblock copolymer induced from microphase separated annealed melt at** $U_p = 0.6$ **for (a)** $x_B = 0.125$, **(b)** $x_B = 0.25$, **(c)** $x_B = 0.375$, **(d)** $x_B = 0.625$, **(e)** $x_B = 0.75$ **and (f)** $x_B = 0.875$. **Blue and orange lines represent crystalline bonds of A-block and B-block respectively, and yellow lines represent non-crystalline bonds of both the blocks.**

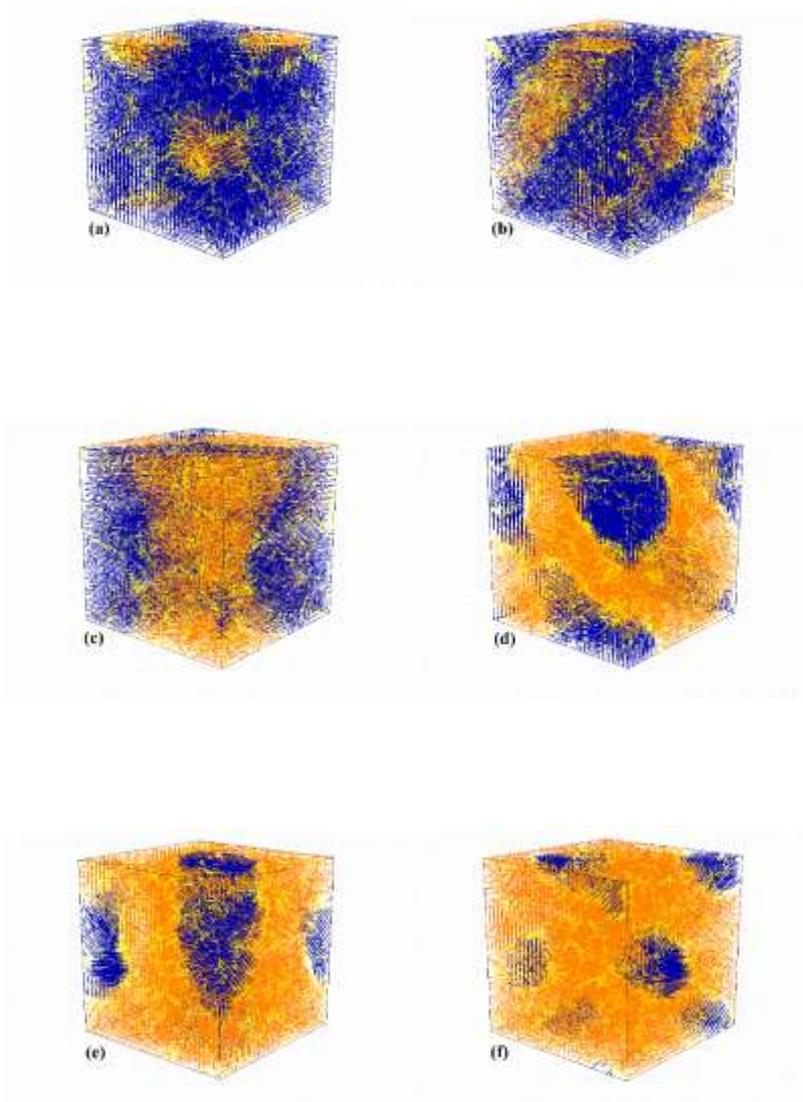

**Figure S11. Change in scaled crystallinity ($X_c^*$) with Monte Carlo Steps (*MCS*) at $U_p$ = 0.6 for (a) A-block and (b) B-block introduced by microphase separate melt without annealing during one-step isothermal crystallization.**

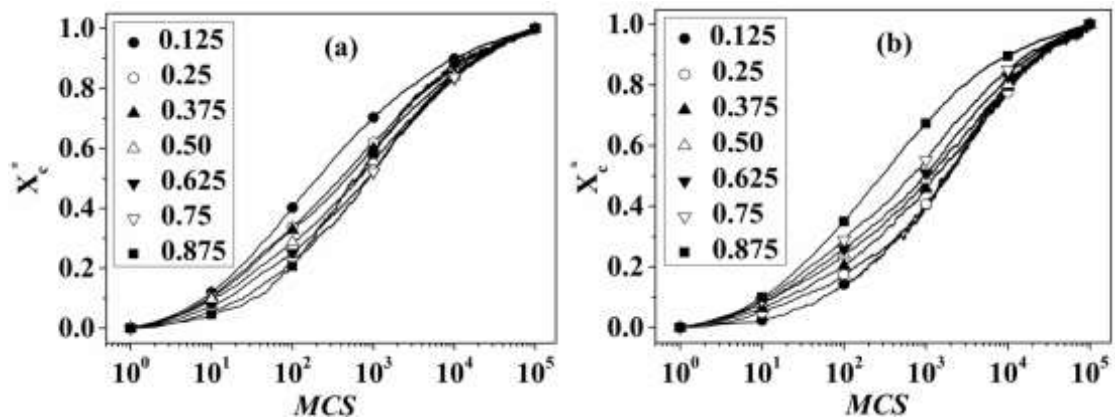

**Figure S12. Change in scaled crystallinity ($X_c^*$) with Monte Carlo Steps (*MCS*) at $U_p$ = 0.6 for (a) A-block and (b) B-block introduced by microphase separate annealed melt during one-step isothermal crystallization.**

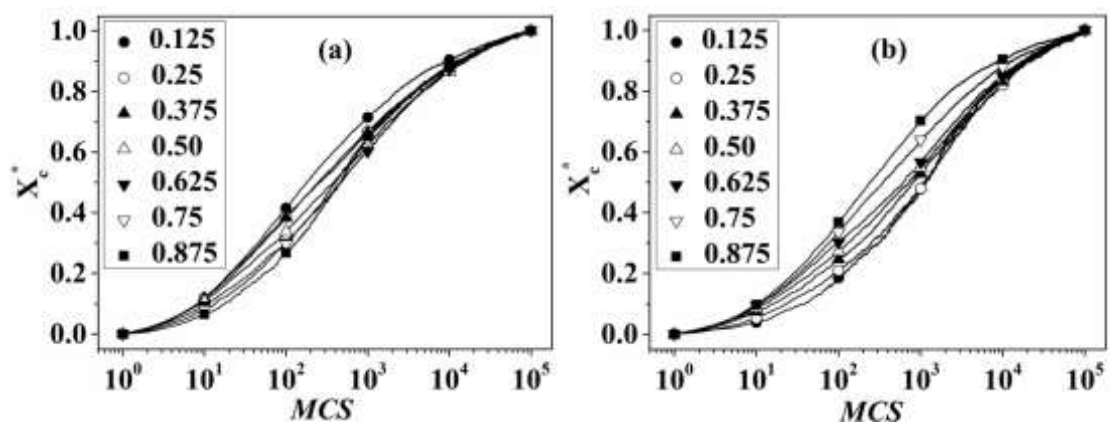

**Figure S13: Change in mean square radius of gyration** $\left\langle R_g^2 \right\rangle$ **with** *MCS* **at** $U_p = 0.6$ **for (a)** $x_B = 0.125$, **(b)** $x_B = 0.25$, **(c)** $x_B = 0.375$, **(d)** $x_B = 0.625$, **(e)** $x_B = 0.75$ **and (f)** $x_B = 0.875$ **from microphase separated melt without and with annealing. There is no substantial change in the value of** $\left\langle R_g^2 \right\rangle$ **with Monte Carlo steps for both the melt systems.**

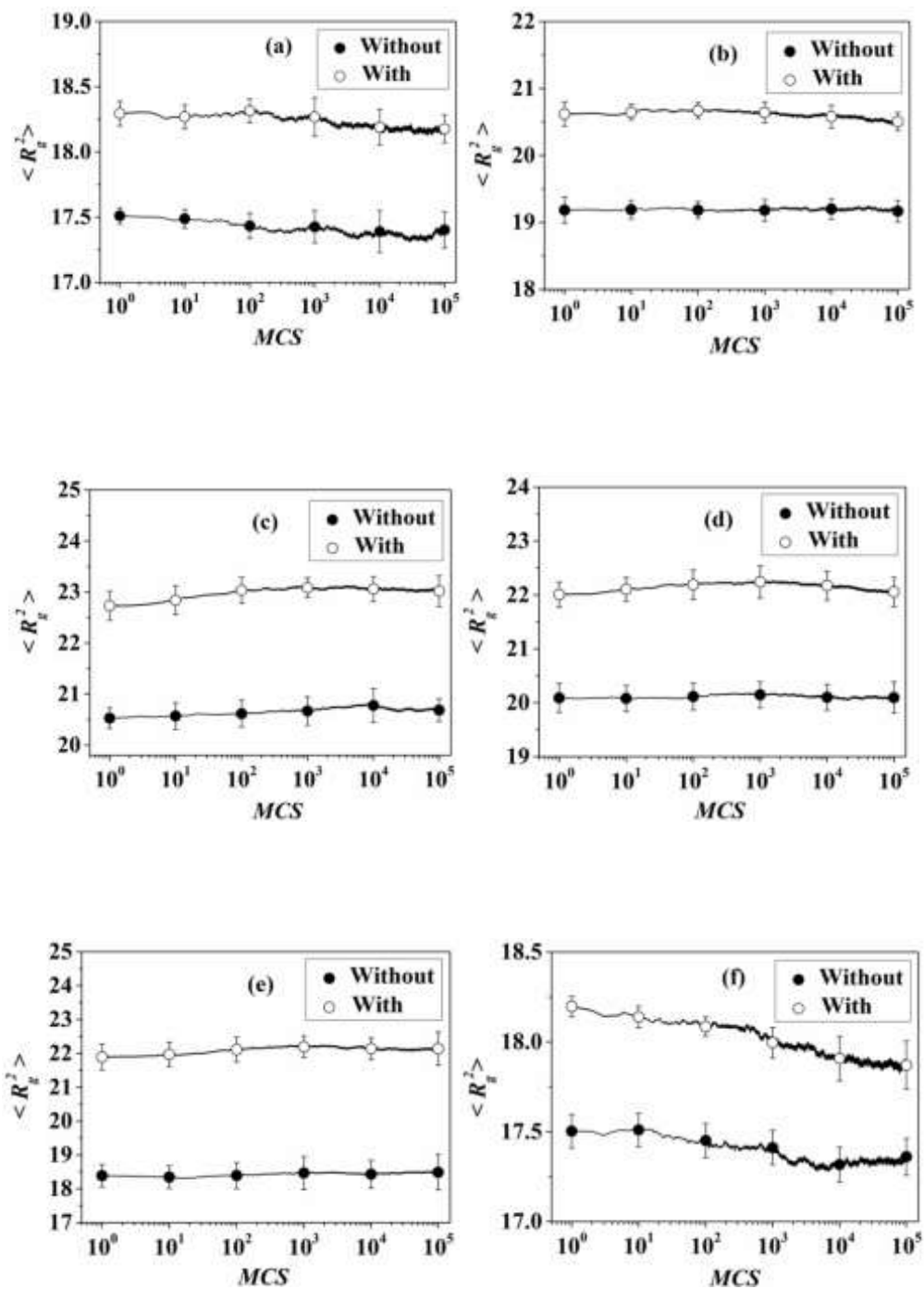

**Figure S14. Snapshots of semi-crystalline structure of diblock copolymer induced from microphase separated melt without annealing at $U_p$ = 0.6 for (a) $x_B$ = 0.125, (b) $x_B$ = 0.25, (c) $x_B$ = 0.375, (d) $x_B$ = 0.625, (e) $x_B$ = 0.75 and (f) $x_B$ = 0.875 after one-step isothermal crystallization. Blue and orange lines represent crystalline bonds of A-block and B-block respectively, and yellow lines represent non-crystalline bonds of both the blocks.**

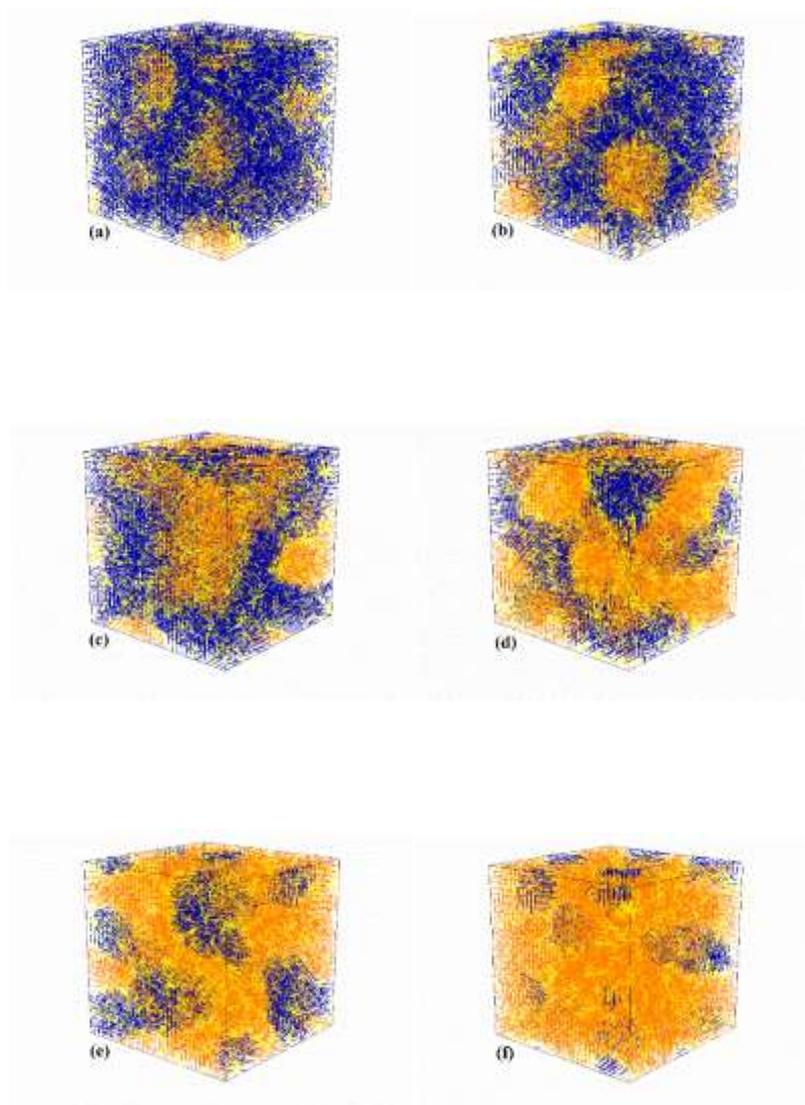

**Figure S15. Snapshots of semi-crystalline structure of diblock copolymer induced from microphase separated annealed melt at** $U_p$ **= 0.6 for (a)** $x_B$ **= 0.125, (b)** $x_B$ **= 0.25, (c)** $x_B$ **= 0.375, (d)** $x_B$ **= 0.625, (e)** $x_B$ **= 0.75 and (f)** $x_B$ **= 0.875 after one-step isothermal crystallization. Blue and orange lines represent crystalline bonds of A-block and B-block respectively, and yellow lines represent non-crystalline bonds of both the blocks.**

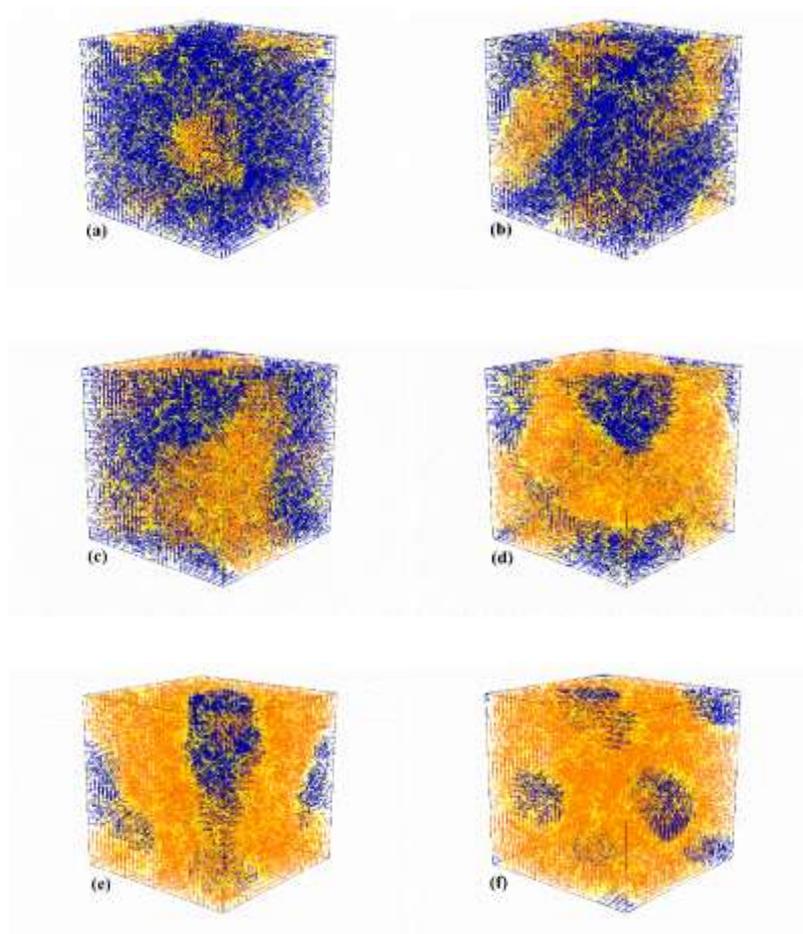

**Table S1. Comparison in lamellar thickness of diblock copolymer induced from microphase separated melt without annealing during two-step and one-step crystallization.**

| $x_B$ | A-Block | | B-block | |
|---|---|---|---|---|
| | Two-step | One-step | Two-step | One-step |
| 0.125 | 3.27 | 2.52 | 2.59 | 2.67 |
| 0.25 | 3.21 | 2.46 | 2.54 | 2.60 |
| 0.375 | 3.23 | 2.46 | 2.54 | 2.55 |
| 0.50 | 3.25 | 2.46 | 2.50 | 2.52 |
| 0.625 | 3.46 | 2.50 | 2.51 | 2.52 |
| 0.75 | 3.69 | 2.59 | 2.55 | 2.55 |
| 0.875 | 3.38 | 2.79 | 2.60 | 2.62 |

**Table S2. Comparison in lamellar thickness of diblock copolymer induced from microphase separated annealed melt during two-step and one-step crystallization.**

| $x_B$ | A-Block | | B-block | |
|---|---|---|---|---|
| | **Two-step** | **One-step** | **Two-step** | **One-step** |
| 0.125 | 3.26 | 2.53 | 2.66 | 2.70 |
| 0.25 | 3.23 | 2.48 | 2.57 | 2.61 |
| 0.375 | 3.25 | 2.48 | 2.55 | 2.58 |
| 0.50 | 3.25 | 2.48 | 2.53 | 2.55 |
| 0.625 | 3.42 | 2.53 | 2.53 | 2.55 |
| 0.75 | 3.60 | 2.63 | 2.60 | 2.59 |
| 0.875 | 4.20 | 2.83 | 2.65 | 2.66 |

**Figure S16. Snapshots of diblock copolymer induced from microphase separated melt without annealing for (a)** $x_B$ **= 0.25 at** $U_p$ **= 0.3 (b)** $x_B$ **= 0.25 at** $U_p$ **= 0.6 (c)** $x_B$ **= 0.75 at** $U_p$ **= 0.3 and (d)** $x_B$ **= 0.75 at** $U_p$ **= 0.6 during two-step isothermal crystallization. Blue and orange lines represent crystalline bonds of A-block and B-block respectively, and yellow lines represent non-crystalline bonds of both the blocks.**

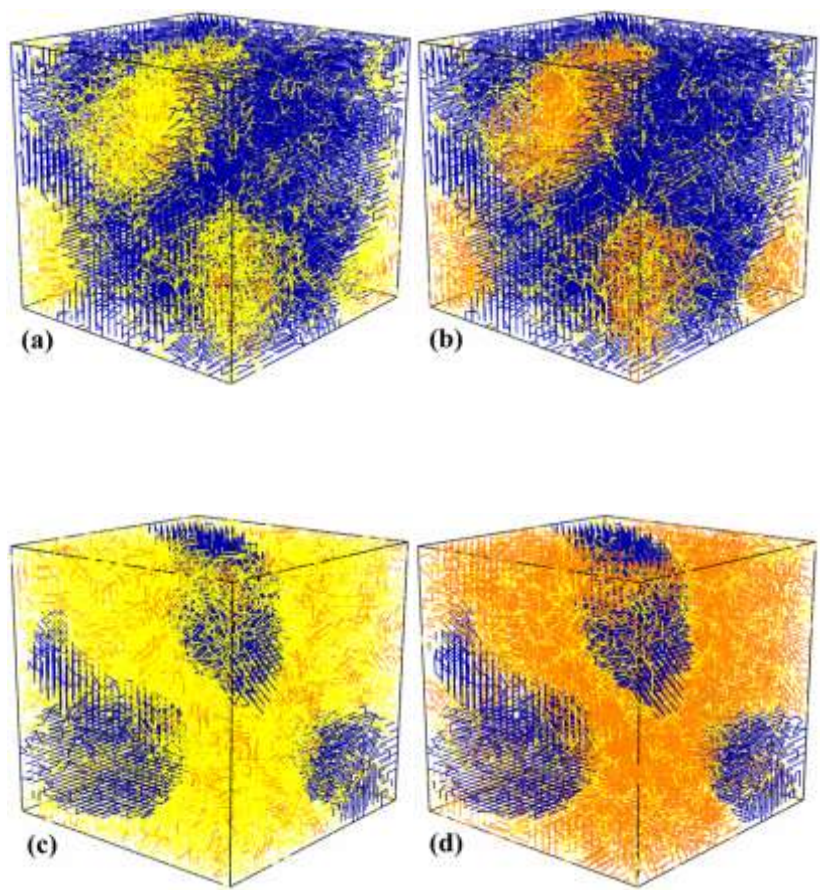

**Figure S17.** Snapshots of diblock copolymer induced from microphase separated annealed melt for **(a)** $x_B$ = **0.25 at** $U_p$ = **0.3 (b)** $x_B$ = **0.25 at** $U_p$ = **0.6 (c)** $x_B$ = **0.75 at** $U_p$ = **0.3 and (d)** $x_B$ = **0.75 at** $U_p$ = **0.6 during two-step isothermal crystallization. Blue and orange lines represent crystalline bonds of A-block and B-block respectively, and yellow lines represent non-crystalline bonds of both the blocks.**

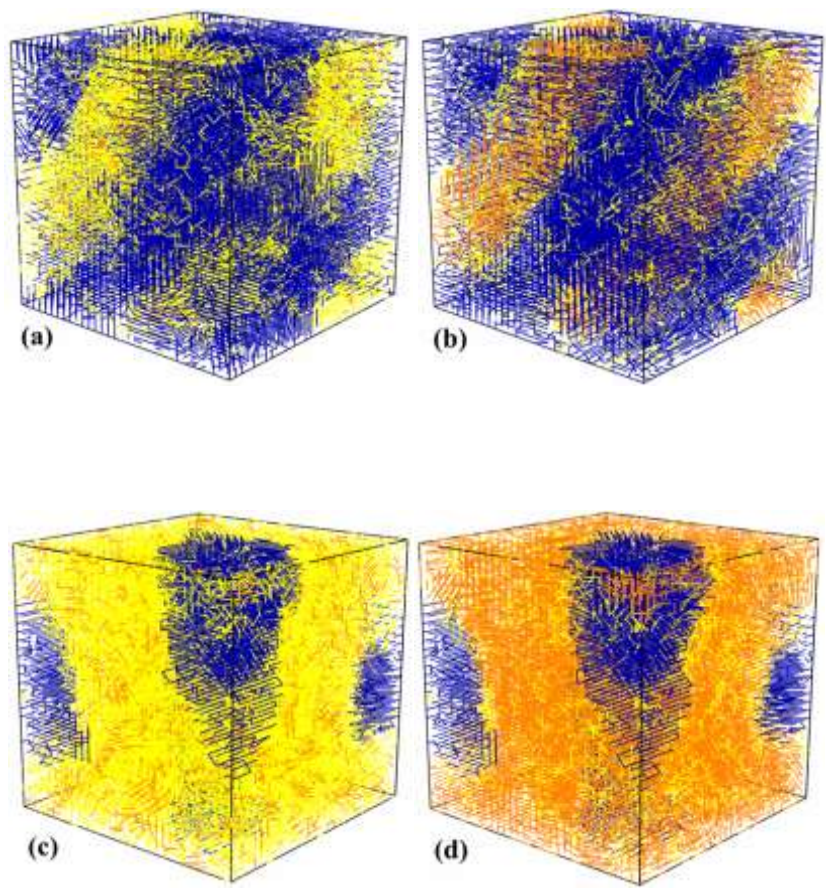